\documentclass[preprint]{ptephy}

\preprintnumber{KYUSHU-HET-162}

\usepackage{amssymb}
\usepackage{amsthm}
\usepackage{amsmath}
\usepackage{booktabs}
\DeclareMathOperator{\tr}{tr}

\newcommand{\Slash}[1]{{\ooalign{\hfil/\hfil\crcr$#1$}}}
\numberwithin{equation}{section}

\usepackage{hyperref}
\usepackage{esint}

\begin{document}

\title{Proof of the renormalizability of the gradient flow}

\author{%
\name{\fname{Kenji} \surname{Hieda}}{1},
\name{\fname{Hiroki} \surname{Makino}}{1}, and
\name{\fname{Hiroshi} \surname{Suzuki}}{1,\ast}
}

\address{%
\affil{1}{Department of Physics, Kyushu University, 744 Motooka, Nishi-ku,
Fukuoka, 819-0395, Japan}
\email{hsuzuki@phys.kyushu-u.ac.jp}
}

\begin{abstract}%
We give an alternative perturbative proof of the renormalizability of the
system defined by the gradient flow and the fermion flow in vector-like gauge
theories.
\end{abstract}
\subjectindex{B01, B31, B32, B38}
\maketitle

\section{Introduction}
\label{sec:1}
The non-abelian gauge theory is renormalizable in the sense that one can send
the momentum cutoff infinity while keeping the strength of the interaction
finite. In such a theory with an infinite cutoff, any observable must be
renormalized one and hence in gauge theory how to construct renormalized
quantities is a fundamental question. The gradient or Wilson
flow~\cite{Luscher:2009eq,Luscher:2010iy} provides a surprisingly versatile
method to define renormalized quantities in gauge theory, without explicitly
referring to the perturbative renormalization. See
also~Ref.~\cite{Narayanan:2006rf}. Since renormalized quantities have the
meaning being independent of regularization, this method is especially useful
in the context of lattice regularization with which one clearly wants to define
renormalized quantities without referring to perturbation theory as much as
possible.

That the bare gauge field and its composite operators evolved by the gradient
flow possess finite correlation functions without any wave function
renormalization was proven in~Ref.~\cite{Luscher:2011bx} to all orders of
perturbation theory. The gradient flow has been generalized to the flow of the
fermion field~\cite{Luscher:2013cpa} and it is known that wave function
renormalization of the flowed fermion elementary field makes any correlation
functions and composite operators finite. Backed by these theoretical
understandings, the gradient flow has been applied to many problems in lattice
gauge theory; Refs.~\cite{Luscher:2013vga} and~\cite{Ramos:2015dla} are recent
reviews. More recently, as the reference list~\cite{%
Ce:2015qha,
Fodor:2015zna,
Datta:2015iea,
Shindler:2015aqa,
Suzuki:2015bqa,
Bellwied:2015lba,
Abdel-Rehim:2015pwa,
Hasenfratz:2015ssa,
Francis:2015daa,
Kagimura:2015via,
Ramos:2015baa,
Bornyakov:2015eaa,
Borsanyi:2015cka,
Fukaya:2015ara,
Alexandrou:2015yba,
Petreczky:2015yta,
Chowdhury:2015hta,
Endres:2015yca,
DeGrand:2015zxa,
Lin:2015zpa,
Alexandrou:2015spa,
Yamamura:2015kva,
Boyle:2015exm,
Grabowska:2015qpk,
Brower:2015owo,
Rantaharju:2015cne,
Datta:2015bzm,
Banerjee:2015qtp,
Mages:2015scv,
Giedt:2015alr,
Fujikawa:2016qis,
Appelquist:2016viq,
Vera:2016xpp,
Boyle:2016imm,
Arthur:2016dir,
Wandelt:2016oym,
Chiu:2016uui}
shows, the number of related works is rapidly growing (see also proceedings of
the recent lattice conferences). In most of these works, the gradient flow is
used to the scale setting, definition of a non-perturbative running coupling,
and to define the topological charge. Another interesting branch of application
is to define a wide class of renormalized composite operators using the flow,
such as the energy--momentum
tensor~\cite{Suzuki:2013gza,Makino:2014taa}\footnote{The validity of these
constructions has been tested
numerically~\cite{Asakawa:2013laa,Kitazawa:2014uxa,Itou:2015gxx,%
Kitazawa:2015ewp,Kitazawa:2016,Taniguchi:2016}.} and the axial-vector
current~\cite{Endo:2015iea}. In Refs.~\cite{DelDebbio:2013zaa,Patella:2014dsa,%
Capponi:2015ucc,Capponi:2015ahp}, applications of the gradient flow for the
construction of the energy--momentum tensor are studied from a somewhat
different perspective
from~Refs.~\cite{Suzuki:2013gza,Makino:2014taa}.\footnote{For recent
developments in the problem of the construction of the energy--momentum tensor
in lattice gauge theory, see Refs.~\cite{Giusti:2014tfa,Giusti:2015daa,%
Pepe:2015kda,Giusti:2015got}.} See also Refs.~\cite{Makino:2014sta,%
Makino:2014cxa,Aoki:2014dxa,Suzuki:2015fka,Monahan:2015lha,Aoki:2015dla,%
Kamata:2015ymx} for related works.

In the present paper, we present an alternative perturbative proof of the
renormalizability of the system defined by the gradient flow and the fermion
flow. Although the proof of the renormalizability has already been given
in~Refs.~\cite{Luscher:2011bx,Luscher:2013cpa}, at some stages of the proof,
the explanations in~Refs.~\cite{Luscher:2011bx,Luscher:2013cpa} seem rather
concise. In the present paper, we tried to improve the presentations in the
following points. As Refs.~\cite{Luscher:2011bx,Luscher:2013cpa}, we use a
$(D+1)$-dimensional field theoretical representation of the system defined by
the flow. However, to show the renormalizability, we employ the Ward--Takahashi
(WT) relation associated with the Becchi--Rouet--Stora (BRS)
symmetry~\cite{Becchi:1975nq} expressed in terms of the effective action (the
generating functional of the 1PI correlation functions), the Zinn-Justin
equation~\cite{ZinnJustin:1974mc,Lee:1975vt} (see Ref~\cite{Taylor:1976ru} for
example for the exposition). This type of proof, that was adopted also
in~Ref.~\cite{Makino:2014sta} to prove the renormalizability of the gradient
flow in the 2D $O(N)$ non-linear sigma model, is easy to follow step by step,
although tends to become somewhat lengthy.\footnote{In our proof, one can find
expressions being analogue to those in the proof of the renormalizability of
the stochastic quantization~\cite{ZinnJustin:1986eq,ZinnJustin:1987ux}.} Also,
in the present paper, as in~Ref.~\cite{Makino:2014sta}, we provide a precise
definition of the flow-time derivative by the forward difference prescription.
We show that this prescription is consistent with the above BRS symmetry on
which the derivation of the WT relation relies. This precise definition
resolves a certain subtle issue concerning the so-called ``flow-line loop'';
the vanishing-ness of the flow-line loop is crucially important for the
renormalizability proof. The prescription also clearly illustrates the origin
of the ``boundary counterterm''. We believe that our proof is quite accessible
because of these aspects.

This paper is organized as follows. In~Sect.~\ref{sec:2}, we recapitulate the
basic definition of the gradient flow and the fermion flow in vector-like gauge
theories. We then elucidate the perturbative expansion of the flowed system.
In~Sect.~\ref{sec:3}, we introduce a $(D+1)$-dimensional local field
theory~\cite{Luscher:2011bx,Luscher:2013cpa} whose Feynman rules reproduce the
perturbative expansion of the flowed system in~Sect.~\ref{sec:2.2}. We lay some
emphasis on the usefulness of discretized flow time; this resolves a certain
subtle issue concerning a ``flow-line loop'', that is crucial for the
perturbative equivalence of the $(D+1)$-dimensional field theory and the flowed
system. Section~\ref{sec:4} is the main part of the present paper, the proof of
the renormalizability. First, we study the case of the pure Yang--Mills theory.
In Sect.~\ref{sec:4.1}, we derive the WT relation in the $(D+1)$-dimensional
field theory. After stating our way of multiplicative renormalization
(Eqs.~\eqref{eq:(4.15)}--\eqref{eq:(4.19)}) and expressing the WT relation in
terms of renormalized quantities (Sect.~\ref{sec:4.2}), we argue the most
general form of the divergent part in the effective action
(Sect.~\ref{sec:4.3}). Finally, by examining the constraint imposed by the WT
relation on the divergent part, we find the validity of our
renormalization~\eqref{eq:(4.15)}--\eqref{eq:(4.19)} (Sect.~\ref{sec:4.4}) to
all orders of perturbation theory. In~Sect.~\ref{sec:4.5}, the proof is
generalized to include the fermion fields. The renormalization factors are
defined by~Eqs.~\eqref{eq:(4.51)}--\eqref{eq:(4.57)}. Here, we find that the
flowed fermion fields require wave function
renormalization~\cite{Luscher:2013cpa}, in contrast to the flowed gauge field.
In Sect.~\ref{sec:4.6}, we argue the finiteness of composite operators of the
flowed fields. Section~\ref{sec:5} is devoted for the conclusion.

Here is our notation: The spacetime dimension is set
to~$D\equiv 4-2\varepsilon$ because in what follows we implicitly assume
dimensional regularization that preserves the gauge symmetry in perturbation
theory. The signature of the metric is euclidean and all Dirac matrices are
hermitian. The summation over repeated Lorentz and gauge indices is understood.
The generators of the gauge group are normalized as
\begin{equation}
   \tr(T^aT^b)=-\frac{1}{2}\delta^{ab}.
\end{equation}
We use the following abbreviation for the momentum integral:
\begin{equation}
   \int_p\equiv\int\frac{d^Dp}{(2\pi)^D}.
\end{equation}

\section{Gradient flow, fermion flow and the perturbative expansion}
\label{sec:2}
\subsection{The $D$-dimensional system}
\label{sec:2.1}
We consider the gradient flow and the fermion flow in the vector-like gauge
theory defined by
\begin{equation}
   S=-\frac{1}{2g_0^2}\int d^Dx\,\tr\left[F_{\mu\nu}(x)F_{\mu\nu}(x)\right]
   +\int d^Dx\,\Bar{\psi}(x)(\Slash{D}+m_0)\psi(x),
\label{eq:(2.1)}
\end{equation}
where $g_0$ and~$m_0$ are the bare gauge coupling and the bare mass parameter,
respectively. The field strength is given by
\begin{equation}
   F_{\mu\nu}(x)
   =\partial_\mu A_\nu(x)-\partial_\nu A_\mu(x)+[A_\mu(x),A_\nu(x)]
\label{eq:(2.2)}
\end{equation}
from the gauge potential $A_\mu(x)=A_\mu^a(x)T^a$ and the covariant derivative
on the fermion field~$\psi(x)$ in~$\Slash{D}=\gamma_\mu D_\mu$ is defined by
\begin{equation}
   D_\mu=\partial_\mu+A_\mu.
\label{eq:(2.3)}
\end{equation}

To define perturbation theory, we introduce the gauge fixing term and the
Faddeev--Popov ghost term by
\begin{equation}
   S_{\text{gf}}+S_{c\Bar{c}}
   \equiv
   \delta\frac{-2}{g_0^2}\int d^Dx\,
   \tr\left\{\Bar{c}(x)\left[\partial_\mu A_\mu(x)-\frac{1}{2\lambda_0}B(x)
   \right]\right\},
\label{eq:(2.4)}
\end{equation}
where $\delta$ is the nilpotent BRS transformation on the gauge field, the
ghost field ($c(x)=c^a(x)T^a$), the anti-ghost field
($\Bar{c}(x)=\Bar{c}^a(x)T^a$), and the Nakanishi--Lautrup field
($B(x)=B^a(x)T^a$):
\begin{align}
   \delta A_\mu(x)&=D_\mu c(x),&
   \delta c(x)&=-c(x)^2,
\label{eq:(2.5)}
\\
   \delta\psi(x)&=-c(x)\psi(x),&\delta\Bar{\psi}(x)&=-\Bar{\psi}(x)c(x),
\label{eq:(2.6)}
\\
   \delta\Bar{c}(x)&=B(x),&
   \delta B(x)&=0.
\label{eq:(2.7)}
\end{align}
From these constructions, the above $D$-dimensional field theory is manifestly
BRS invariant:
\begin{equation}
   \delta S=0,\qquad\delta\left(S_{\text{gf}}+S_{c\Bar{c}}\right)=0.
\label{eq:(2.8)}
\end{equation}

\subsection{Flow equations}
\label{sec:2.2}
The Yang--Mills gradient flow is an evolution of a gauge field
configuration~$A_\mu(x)$ along a fictitious time (the flow time)
$t\in[0,\infty)$, according to the flow
equation~\cite{Luscher:2009eq,Luscher:2010iy}
\begin{equation}
   \partial_tB_\mu(t,x)=D_\nu G_{\nu\mu}(t,x)
   +\alpha_0D_\mu\partial_\nu B_\nu(t,x),\qquad
   B_\mu(t=0,x)=A_\mu(x),
\label{eq:(2.9)}
\end{equation}
where $G_{\mu\nu}(t,x)$ is the field strength of the flowed field,
\begin{equation}
   G_{\mu\nu}(t,x)
   =\partial_\mu B_\nu(t,x)-\partial_\nu B_\mu(t,x)
   +[B_\mu(t,x),B_\nu(t,x)],
\label{eq:(2.10)}
\end{equation}
and $D_\mu$ is the covariant derivative on the gauge field,
\begin{equation}
   D_\mu=\partial_\mu+[B_\mu,\cdot].
\label{eq:(2.11)}
\end{equation}
The first term on the right-hand side of the flow equation~\eqref{eq:(2.9)} is
the ``gradient'' in the functional space, $-g_0^2\delta S/\delta B_\mu(t,x)$,
where $S$~is the Yang--Mills action integral for the flowed field. The flow
equation~\eqref{eq:(2.9)} has a form of the diffusion equation and the
evolution for~$t>0$ effectively damps high-frequency modes in the
configuration. The second term on the right-hand side of the flow
equation~\eqref{eq:(2.9)} is a ``gauge fixing term'' and it makes the integrand
of the Feynman integral well behaved. It can be shown~\cite{Luscher:2010iy}
that, however, any gauge-invariant quantity which does not contain the
flow-time derivative~$\partial_t$ is independent of the ``gauge fixing
parameter''~$\alpha_0$.

An evolution of fermion fields being similar to the above can be introduced. A
possible choice is~\cite{Luscher:2013cpa}
\begin{align}
   &\partial_t\chi(t,x)=\left[\Delta-\alpha_0\partial_\mu B_\mu(t,x)\right]
   \chi(t,x),&
   \chi(t=0,x)&=\psi(x),
\label{eq:(2.12)}\\
   &\partial_t\Bar{\chi}(t,x)
   =\Bar{\chi}(t,x)
   \left[\overleftarrow{\Delta}
   +\alpha_0\partial_\mu B_\mu(t,x)\right],
   &\Bar{\chi}(t=0,x)&=\Bar{\psi}(x),
\label{eq:(2.13)}
\end{align}
where
\begin{align}
   \Delta&=D_\mu D_\mu,&D_\mu&=\partial_\mu+B_\mu,
\label{eq:(2.14)}\\
   \overleftarrow{\Delta}&=\overleftarrow{D}_\mu\overleftarrow{D}_\mu,&
   \overleftarrow{D}_\mu&\equiv\overleftarrow{\partial}_\mu-B_\mu.
\label{eq:(2.15)}
\end{align}
These evolutions of the fermion fields are referred to as the fermion flow in
what follows.

\subsection{Perturbative solutions to the flow equations}
\label{sec:2.3}
The flow equation for the gauge field, Eq.~\eqref{eq:(2.9)}, can be formally
solved as~\cite{Luscher:2010iy}
\begin{equation}
   B_\mu(t,x)
   =\int d^Dy\left[
   K_t(x-y)_{\mu\nu}A_\nu(y)
   +\int_0^tds\,K_{t-s}(x-y)_{\mu\nu}R_\nu(s,y)
   \right],
\label{eq:(2.16)}
\end{equation}
where
\begin{equation}
   K_t(x)_{\mu\nu}\equiv\int_p\frac{e^{ipx}}{p^2}
   \left[(\delta_{\mu\nu}p^2-p_\mu p_\nu)e^{-tp^2}
   +p_\mu p_\nu e^{-\alpha_0tp^2}\right]
\label{eq:(2.17)}
\end{equation}
is termed the \emph{heat kernel\/} in what follows, and
\begin{equation}
   R_\mu=2[B_\nu,\partial_\nu B_\mu]
   -[B_\nu,\partial_\mu B_\nu]
   +(\alpha_0-1)[B_\mu,\partial_\nu B_\nu]
   +[B_\nu,[B_\nu,B_\mu]]
\label{eq:(2.18)}
\end{equation}
arises from the non-linear terms in the flow equation. Then by iteratively
solving Eq.~\eqref{eq:(2.16)}, we have a perturbative expansion for the flowed
field~$B_\mu(t,x)$ in terms of the initial value~$A_\nu(y)$. Note the retarded
nature of the solution~\eqref{eq:(2.16)}; the heat kernels connect a
sequence of points, $(t,x)$, $(s_1,y_1)$, $(s_2,y_2)$, \dots, in the temporal
order $t\geq s_1\geq s_2\geq s_3\geq \dotsb \geq0$. Note also that the Gaussian
damping factor in the heat kernel~$K_{t-s}(x)_{\mu\nu}$ behaves
as~$\sim e^{-(t-s)p^2}$ or~$\sim e^{-\alpha_0(t-s)p^2}$, where $t$ and~$s$ are flow
times at end points of the heat kernel, so the damping factors become inactive
when the two flow times $t$ and~$s$ coincide.

For the fermion flow~\eqref{eq:(2.12)} and~\eqref{eq:(2.13)}, we have
\begin{align}
   \chi(t,x)&=\int d^Dy\,
   \left[
   K_t(x-y)\psi(y)
   +\int_0^tds\,K_{t-s}(x-y)\Delta'\chi(s,y)
   \right],
\label{eq:(2.19)}\\
   \Bar{\chi}(t,x)&=\int d^Dy\,
   \left[
   \Bar{\psi}(y)K_t(x-y)
   +\int_0^tds\,\Bar{\chi}(s,y)\overleftarrow{\Delta}'K_{t-s}(x-y)
   \right],
\label{eq:(2.20)}
\end{align}
where
\begin{equation}
   K_t(x)\equiv\int_pe^{ipx}e^{-tp^2}
\label{eq:(2.21)}
\end{equation}
is also referred to as the \emph{heat kernel}, and
\begin{align}
   \Delta'&\equiv(1-\alpha_0)\partial_\mu B_\mu
   +2B_\mu\partial_\mu+B_\mu B_\mu,
\label{eq:(2.22)}\\
   \overleftarrow{\Delta}'&
   \equiv-(1-\alpha_0)\partial_\mu B_\mu
   -2\overleftarrow{\partial}_\mu B_\mu+B_\mu B_\mu.
\label{eq:(2.23)}
\end{align}
By iteratively solving Eqs.~\eqref{eq:(2.19)} and~\eqref{eq:(2.20)}, we again
have a perturbative expansion of the fermion flow. Again note the retarded
nature of the solutions and the fact that the Gaussian factor behaves
as~$\sim e^{-(t-s)p^2}$.

The initial values for the above flows, $A_\mu(x)$, $\psi(x)$,
and~$\Bar{\psi}(x)$, are quantum fields being subject to the functional
integral. Correlation functions of the flowed fields are thus obtained by
expressing the flowed fields in terms of the initial values following the
above procedures and then carrying out the functional integral over the latter.
For example, in the tree-level approximation, we have
\begin{equation}
   \left\langle B_\mu^a(t,x)B_\nu^b(s,y)\right\rangle_0
   =\delta^{ab}g_0^2\int_p
   \frac{e^{ip(x-y)}}{(p^2)^2}
   \left[
   (\delta_{\mu\nu}p^2-p_\mu p_\nu)e^{-(t+s)p^2}
   +\frac{1}{\lambda_0}p_\mu p_\nu e^{-\alpha_0(t+s)p^2}
   \right]
\label{eq:(2.24)}
\end{equation}
and, similarly for the fermion field,
\begin{equation}
   \left\langle\chi(t,x)\Bar{\chi}(s,y)\right\rangle_0
   =\int_p
   \frac{e^{ip(x-y)}}{i\Slash{p}+m_0}e^{-(t+s)p^2},
\label{eq:(2.25)}
\end{equation}
in the tree-level approximation. Eqs.~\eqref{eq:(2.24)} and~\eqref{eq:(2.25)}
are referred to as \emph{propagators\/} in what follows.

In a diagrammatic representation of the above perturbative expansion of the
flowed system (see Refs.~\cite{Luscher:2010iy,Luscher:2011bx,Makino:2014taa}
for details), we thus have two types of ``lines'', one represents the heat
kernels~\eqref{eq:(2.17)} and~\eqref{eq:(2.21)}, and another is for the
propagators~\eqref{eq:(2.24)} and~\eqref{eq:(2.25)}. The former line is called
the \emph{flow line} (or heat kernel), while the latter is simply called the
propagator. Note that the propagators carry the Gaussian damping factor of the
form $\sim e^{-(t+s)p^2}$ or~$\sim e^{-\alpha_0(t+s)p^2}$, where $t$ and~$s$ are
flow times at end points of the propagator, so the damping factor is always
active even when the two flow times $t$ and~$s$ coincide as far as $t>0$
or~$s>0$; this is a crucial difference from the heat kernel or the flow line.
Also we have two types of vertices, one arises from the non-linear terms in the
flow equations, Eqs.~\eqref{eq:(2.18)}, \eqref{eq:(2.22)},
and~\eqref{eq:(2.23)}, and another represents the conventional interaction
vertices in~Eq.~\eqref{eq:(2.1)}. An important feature of this diagrammatic
representation of the flowed system is that there is no closed loop totally
consisting of the flow lines (or heat kernels). This property, which turns out
to be crucial for the renormalizability of the flowed system, follows from the
fact that to form a loop in a diagram, we have to Wick contract the initial
values and then the contraction results in one of the
propagators~\eqref{eq:(2.24)} and~\eqref{eq:(2.25)}, but not heat kernels.

The statement of the renormalizability of the flowed system, first proved
in~Refs.~\cite{Luscher:2011bx,Luscher:2013cpa} to all orders of perturbation
theory, is that any correlation functions and composite operators of flowed
fields are made finite by the conventional renormalization in the gauge theory
and wave function renormalization of the flowed fermion fields. In particular,
no wave function renormalization of the flowed gauge field is required. The
purpose of the present paper is to give an another proof of this statement.

\section{A $(D+1)$-dimensional field theoretical representation of the
flowed system}
\label{sec:3}
\subsection{$(D+1)$-dimensional field theory and the BRS invariance}
\label{sec:3.1}
The total action of a $(D+1)$-dimensional local field theory we will consider
is
\begin{equation}
   S_{\text{tot}}=S+S_{\text{gf}}+S_{c\Bar{c}}+S_{\text{fl}}+S_{d\Bar{d}},
\label{eq:(3.1)}
\end{equation}
where $D$-dimensional actions, $S$, $S_{\text{gf}}$, and~$S_{c\Bar{c}}$, are
already introduced in~Sect.~\ref{sec:2.1}. The $(D+1)$-dimensional actions are
defined by~\cite{Luscher:2011bx,Luscher:2013cpa}
\begin{align}
   S_{\text{fl}}
   &=-2\int_0^\infty dt\int d^Dx\,
   \tr\left\{
   L_\mu(t,x)\left[
   \partial_tB_\mu(t,x)-D_\nu G_{\nu\mu}(t,x)
   -\alpha_0D_\mu\partial_\nu B_\nu(t,x)
   \right]
   \right\}
\notag\\
   &\qquad{}
   +\int_0^\infty dt\,\int d^Dx\,
   \Bigl\{\Bar{\lambda}(t,x)
   \left[\partial_t-\Delta+\alpha_0\partial_\mu B_\mu(t,x)\right]\chi(t,x)
\notag\\
   &\qquad\qquad\qquad\qquad\qquad{}+
   \Bar{\chi}(t,x)
   \left[\overleftarrow{\partial}_t
   -\overleftarrow{\Delta}-\alpha_0\partial_\mu B_\mu(t,x)\right]
   \lambda(t,x)
   \Bigr\}
\label{eq:(3.2)}
\end{align}
and
\begin{equation}
   S_{d\Bar{d}}
   =-2\int_0^\infty dt\int d^Dx\,
   \tr\left\{
   \Bar{d}(t,x)\left[
   \partial_td(t,x)-\alpha_0D_\mu\partial_\mu d(t,x)\right]\right\}.
\label{eq:(3.3)}
\end{equation}

In Eqs.~\eqref{eq:(3.1)} and~\eqref{eq:(3.2)}, $(D+1)$-dimensional fields,
$B_\mu(t,x)$, $\chi(t,x)$, and~$\Bar{\chi}(t,x)$, are subject to the boundary
conditions,
\begin{equation}
   B_\mu(t=0,x)=A_\mu(x),\qquad\chi(t=0,x)=\psi(x),
   \qquad\Bar{\chi}(t=0,x)=\Bar{\psi}(x),
\label{eq:(3.4)}
\end{equation}
corresponding to the initial conditions of the flow equations
in~Eqs.~\eqref{eq:(2.9)}, \eqref{eq:(2.12)}, and~\eqref{eq:(2.13)}. The
Lagrange multiplier fields, $L_\mu(t,x)=L_\mu^a(t,x)T^a$, $\lambda(t,x)$,
and~$\Bar{\lambda}(t,x)$, on the other hand, obey free boundary conditions,
that is, $L_\mu(t=0,x)$, $\lambda(t=0,x)$, and~$\Bar{\lambda}(t=0,x)$ are
integrated as independent variables in the functional integral.

In~Eq.~\eqref{eq:(3.3)}, $d(t,x)$ is a $(D+1)$-dimensional ghost field which
obeys the boundary condition,
\begin{equation}
   d(t=0,x)=c(x),
\label{eq:(3.5)}
\end{equation}
and $\Bar{d}(t,x)=\Bar{d}^a(t,x)T^a$ is a ghost analogue of the Lagrange
multiplier field; it also obeys the free boundary condition at~$t=0$.

In addition to~Eqs.~\eqref{eq:(2.5)}--\eqref{eq:(2.7)}, we introduce the BRS
transformation on the above $(D+1)$-dimensional fields by
\begin{align}
   \delta B_\mu(t,x)&=D_\mu d(t,x),&\delta d(t,x)&=-d(t,x)^2,
\label{eq:(3.6)}
\\
   \delta L_\mu(t,x)&=[L_\mu(t,x),d(t,x)],&&
\label{eq:(3.7)}
\\
   \delta\chi(t,x)&=-d(t,x)\chi(t,x),&
   \delta\Bar{\chi}(t,x)&=-\Bar{\chi}(t,x)d(t,x),
\label{eq:(3.8)}
\\
   \delta\lambda(t,x)&=-d(t,x)\lambda(t,x),&
   \delta\Bar{\lambda}(t,x)&=-\Bar{\lambda}(t,x)d(t,x).
\label{eq:(3.9)}
\end{align}
These are simply a standard form of the BRS transformation in which the
parameter of the gauge transformation is replaced by the $(D+1)$-dimensional
ghost field~$d(t,x)$; here $L_\mu(t,x)$ is regarded as a field in the adjoint
representation and $\lambda(t,x)$ ($\Bar{\lambda}(t,x)$) is regarded as
a field in the same representation as $\chi(t,x)$ ($\Bar{\chi}(t,x)$). By
this construction, the nilpotency~$\delta^2=0$ on these fields immediately
follows as for the conventional $D$-dimensional BRS transformation.

On the other hand, for the ``ghost Lagrange multiplier''~$\Bar{d}(t,x)$, we
define the BRS transformation as
\begin{equation}
   \delta\Bar{d}(t,x)
   =D_\mu L_\mu(t,x)-\left\{d(t,x),\Bar{d}(t,x)\right\}
   +\Bar{\lambda}(t,x)T^a\chi(t,x)T^a
   -\Bar{\chi}(t,x)T^a\lambda(t,x)T^a.
\label{eq:(3.10)}
\end{equation}
By a somewhat troublesome calculation, one can confirm that this transformation
on~$\Bar{d}(t,x)$ is nilpotent $\delta^2=0$ without using any equation of
motion. One may wonder how this intricate structure of
nilpotent~$\delta\Bar{d}(t,x)$ arises. In~Appendix~\ref{app:A}, we show that
this structure of~$\delta\Bar{d}(t,x)$ can be naturally understood if one
considers an enlarged field space in which the covariance under a
$(D+1)$-dimensional gauge transformation is manifest.

One can also confirm by a direct calculation that the $(D+1)$-dimensional
system is BRS invariant:
\begin{equation}
   \delta\left(S_{\text{fl}}+S_{d\Bar{d}}\right)=0.
\label{eq:(3.11)}
\end{equation}
This can be seen by writing
\begin{align}
   S_{\text{fl}}+S_{d\Bar{d}}
   &=-2\int_0^\infty dt\int d^Dx\,
   \tr L_\mu(t,x)E_\mu(t,x)
\notag\\
   &\qquad{}
   +\int_0^\infty dt\int d^Dx\,\left[
   \Bar{\lambda}(t,x)f(t,x)
   +\Bar{f}(t,x)\lambda(t,x)\right]
\notag\\
   &\qquad\qquad{}
   -2\int_0^\infty dt\int d^Dx\,
   \tr\Bar{d}(t,x)e(t,x),
\label{eq:(3.12)}
\end{align}
where
\begin{align}
   E_\mu(t,x)&\equiv
   \partial_tB_\mu(t,x)-D_\nu G_{\nu\mu}(t,x)
   -\alpha_0D_\mu\partial_\nu B_\nu(t,x),
\label{eq:(3.13)}
\\
   e(t,x)&\equiv
   \partial_td(t,x)-\alpha_0D_\mu\partial_\mu d(t,x),
\label{eq:(3.14)}
\\
   f(t,x)&\equiv
   \left[\partial_t-\Delta+\alpha_0\partial_\mu B_\mu(t,x)\right]\chi(t,x),
\label{eq:(3.15)}
\\
   \Bar{f}(t,x)&\equiv
   \Bar{\chi}(t,x)
   \left[\overleftarrow{\partial}_t
   -\overleftarrow{\Delta}-\alpha_0\partial_\mu B_\mu(t,x)\right],
\label{eq:(3.16)}
\end{align}
and then using the relations~\cite{Luscher:2011bx}
\begin{align}
   \delta E_\mu(t,x)&=\left[E_\mu(t,x),d(t,x)\right]+D_\mu e(t,x),
\label{eq:(3.17)}
\\
   \delta e(t,x)&=-\left\{e(t,x),d(t,x)\right\},
\label{eq:(3.18)}
\\
   \delta f(t,x)&=-d(t,x)f(t,x)-e(t,x)\chi(t,x),
\label{eq:(3.19)}
\\
   \delta\Bar{f}(t,x)&=-\Bar{f}(t,x)d(t,x)-\Bar{\chi}(t,x)e(t,x).
\label{eq:(3.20)}
\end{align}
Again, this (apparently miraculous) BRS invariance of the action can be
naturally understood in the enlarged field space (see Appendix~\ref{app:A}).

\subsection{Derivation of the Feynman rules}
\label{sec:3.2}
In this subsection, we show that perturbation theory in the above
$(D+1)$-dimensional field theory precisely reproduces the perturbative
expansion of the gradient flow we have considered in~Sect.~\ref{sec:2.3}. This
fact then allows us to use the $(D+1)$-dimensional field theory to prove the
perturbative renormalizability of the flowed system.

For this, it is quite helpful to specify the precise meaning of the flow-time
derivative~$\partial_t$. In what follows, we understand that the flow time is
discretized in the step~$\epsilon$. The integral is thus given by the sum,
\begin{equation}
   \int_0^\infty dt\equiv\epsilon\sum_{t=0}^\infty.
\label{eq:(3.21)}
\end{equation}
For the flow-time derivative, we adopt the \emph{forward difference
prescription},
\begin{equation}
   \partial_t f(t,x)
   \equiv\frac{1}{\epsilon}\left[f(t+\epsilon,x)-f(t,x)\right]
\label{eq:(3.22)}
\end{equation}
for any field~$f(t,x)$. In particular, this prescription implies that
for~$t=0$,
\begin{align}
   \partial_t B_\mu(t=0,x)
   &\equiv\frac{1}{\epsilon}\left[B_\mu(t=\epsilon,x)-A_\mu(x)\right],&&
\label{eq:(3.23)}
\\
   \partial_t\chi(t=0,x)
   &\equiv\frac{1}{\epsilon}\left[\chi(t=\epsilon,x)-\psi(x)\right],&
   \Bar{\chi}(t=0,x)\overleftarrow{\partial}_t
   &\equiv\frac{1}{\epsilon}
   \left[\Bar{\chi}(t=\epsilon,x)-\Bar{\psi}(x)\right],
\label{eq:(3.24)}
\\
   \partial_t d(t=0,x)
   &\equiv\frac{1}{\epsilon}\left[d(t=\epsilon,x)-c(x)\right],&&
\label{eq:(3.25)}
\end{align}
because of the boundary conditions~\eqref{eq:(3.4)} and~\eqref{eq:(3.5)}.

The ordering of removing regularizations is also quite important: After
applying the Feynman rules to a given set of diagrams, we understand that we
first take the continuous flow-time limit~$\epsilon\to0$ and then, after the
renormalization, remove the dimensional regularization, $D\to4$ (or other
$4$-dimensional regularization such as the lattice).

Under the above prescription, to obtain the tree-level propagators, we
introduce the heat kernels with the discretized flow time~\cite{Makino:2014sta}
by
\begin{align}
   K_t^\epsilon(x)_{\mu\nu}
   &\equiv\int_p
   \frac{e^{ipx}}{p^2}
   \left[(\delta_{\mu\nu}p^2-p_\mu p_\nu)(1-\epsilon p^2)^{t/\epsilon}
   +p_\mu p_\nu(1-\epsilon\alpha_0p^2)^{t/\epsilon}
   \right],
\label{eq:(3.26)}
\\
   K_t^\epsilon(x;\alpha_0)
   &\equiv\int_p
   e^{ipx}
   (1-\epsilon\alpha_0p^2)^{t/\epsilon}.
\label{eq:(3.27)}
\end{align}
These quantities satisfy, by construction,
\begin{align}
   \frac{1}{\epsilon}\left[K_{t+\epsilon}^\epsilon(x)_{\mu\nu}
   -K_t^\epsilon(x)_{\mu\nu}\right]
   &=\partial_\rho\partial_\rho K_t^\epsilon(x)_{\mu\nu}
   +(\alpha_0-1)\partial_\mu\partial_\rho K_t^\epsilon(x)_{\rho\nu},
\label{eq:(3.28)}
\\
   K_0^\epsilon(x)_{\mu\nu}&=\delta_{\mu\nu}\delta(x),
\label{eq:(3.29)}
\end{align}
and
\begin{align}
   \frac{1}{\epsilon}\left[K_{t+\epsilon}^\epsilon(x;\alpha_0)
   -K_t^\epsilon(x;\alpha_0)\right]
   &=\alpha_0\partial_\mu\partial_\mu K_t^\epsilon(x;\alpha_0),
\label{eq:(3.30)}
\\
   K_0^\epsilon(x;\alpha_0)&=\delta(x).
\label{eq:(3.31)}
\end{align}

First, we consider the propagators containing the flowed gauge
field~$B_\mu(t,x)$. We change the variable from $B_\mu$ to~$b_\mu$
as~\cite{Luscher:2011bx}
\begin{equation}
   B_\mu(t,x)=\int d^Dy\,K_t^\epsilon(x-y)_{\mu\nu}A_\nu(y)+b_\mu(t,x).
\label{eq:(3.32)}
\end{equation}
Then, because of~Eq.~\eqref{eq:(3.29)}, the boundary condition~\eqref{eq:(3.4)}
becomes simply
\begin{equation}
   b_\mu(t=0,x)=0
\label{eq:(3.33)}
\end{equation}
and, because of~Eq.~\eqref{eq:(3.28)}, the kinetic term of~$b_\mu$ is given by
\begin{align}
   S_{\text{fl}}
   &=\epsilon\sum_{t=0}^\infty\int d^Dx\,
   L_\mu^a(t,x)
\notag\\
   &\qquad{}\times\biggl\{
   \frac{1}{\epsilon}\left[b_\mu^a(t+\epsilon,x)-b_\mu^a(t,x)\right]
   -\partial_\nu\partial_\nu b_\mu^a(t,x)
   +(1-\alpha_0)\partial_\mu\partial_\nu b_\nu^a(t,x)\biggr\}+\dotsb.
\label{eq:(3.34)}
\end{align}
The corresponding Schwinger--Dyson equation in the tree-level yields,
\begin{align}
   &\left\langle\frac{1}{\epsilon}
   \left[b_\mu^a(t+\epsilon,x)-b_\mu^a(t,x)\right]
   L_\nu^b(s,y)
   +\left[
   -\delta_{\mu\rho}\partial_\sigma\partial_\sigma
   +(1-\alpha_0)\partial_\mu\partial_\rho
   \right]b_\rho^a(t,x)L_\nu^b(s,y)\right\rangle_0
\notag\\
   &=\frac{1}{\epsilon}\delta^{ab}\delta_{\mu\nu}\delta_{t,s}\delta(x-y).
\label{eq:(3.35)}
\end{align}
This equation can then be solved step by step in the discretized~$t$, starting
from~$t=0$ at which
\begin{equation}
   \left\langle b_\mu^a(t=0,x)L_\nu^b(s,y)\right\rangle_0=0,\qquad
   s\geq0,
\label{eq:(3.36)}
\end{equation}
because of the boundary condition~\eqref{eq:(3.33)}. In particular, we find
\begin{align}
   \left\langle b_\mu^a(s,x)L_\nu^b(s,y)\right\rangle_0&=0,
\label{eq:(3.37)}
\\
   \left\langle b_\mu^a(s+\epsilon,x)L_\nu^b(s,y)\right\rangle_0
   &=\delta^{ab}\delta_{\mu\nu}\delta(x-y),
\label{eq:(3.38)}
\end{align}
and generally,
\begin{equation}
   \left\langle b_\mu^a(t,x)L_\nu^b(s,y)\right\rangle_0
   =\delta^{ab}\vartheta(t-s)K_{t-s-\epsilon}^\epsilon(x-y)_{\mu\nu},
\label{eq:(3.39)}
\end{equation}
where $\vartheta(t)$ is a ``regularized step function'',
\begin{equation}
   \vartheta(t)\equiv
   \begin{cases}
   1,&\text{for $t>0$},\\
   0,&\text{for $t=0$},\\
   0,&\text{for $t<0$}.\\
   \end{cases}
\label{eq:(3.40)}
\end{equation}

We then note that, from the structure of the action, the $AL$ and $bb$
propagators in the tree-level vanish. Thus, by using~Eq.~\eqref{eq:(3.32)}, we
find
\begin{align}
   &\left\langle B_\mu^a(t,x)L_\nu^b(s,y)\right\rangle_0
   =\delta^{ab}\vartheta(t-s)K_{t-s-\epsilon}^\epsilon(x-y)_{\mu\nu},
\label{eq:(3.41)}
\\
   &\left\langle B_\mu^a(t,x)B_\nu^b(s,y)\right\rangle_0
   =g_0^2\delta^{ab}
   \int_p\frac{e^{ip(x-y)}}{(p^2)^2}
\notag\\
   &\qquad{}\times
   \left[(\delta_{\mu\nu}p^2-p_\mu p_\nu)(1-\epsilon p^2)^{(t+s)/\epsilon}
   +\frac{1}{\lambda_0}p_\mu p_\nu(1-\epsilon\alpha_0p^2)^{(t+s)/\epsilon}\right].
\label{eq:(3.42)}
\end{align}
These results show that, in the $\epsilon\to0$ limit, the former
$BL$~propagator becomes the heat kernel (or flow line) in the perturbative
expansion of the gradient flow, Eq.~\eqref{eq:(2.17)}, and the latter
$BB$~propagator reproduces Eq.~\eqref{eq:(2.24)}.

In a similar way, we find for other flowed fields,
\begin{align}
   \left\langle\chi(t,x)\Bar{\lambda}(s,y)\right\rangle_0
   &=\vartheta(t-s)K_{t-s-\epsilon}^\epsilon(x-y;1),
\label{eq:(3.43)}
\\
   \left\langle\lambda(t,x)\Bar{\chi}(s,y)\right\rangle_0
   &=\vartheta(s-t)K_{s-t-\epsilon}^\epsilon(x-y;1),
\label{eq:(3.44)}
\\
   \left\langle\chi(t,x)\Bar{\chi}(s,y)\right\rangle_0
   &=\int_p\frac{e^{ip(x-y)}}{i\Slash{p}+m_0}(1-\epsilon p^2)^{(t+s)/\epsilon},
\label{eq:(3.45)}
\\
   \left\langle d^a(t,x)\Bar{d}^b(s,y)\right\rangle_0
   &=\delta^{ab}\vartheta(t-s)K_{t-s-\epsilon}^\epsilon(x-y;\alpha_0),
\label{eq:(3.46)}
\end{align}
which, for~$\epsilon\to0$, reproduce heat kernels and propagators
in~Sect.~\ref{sec:2.3}.

From the structure of~$S_{\text{fl}}$ in~Eq.~\eqref{eq:(3.2)}, it is obvious
that the interaction vertices derived from~Eq.~\eqref{eq:(3.2)} are identical
to the non-linear terms in the flow equations, $R_\mu$ in~Eq.~\eqref{eq:(2.18)}
and $\Delta'$ and~$\overleftarrow{\Delta}'$ in~Eqs.~\eqref{eq:(2.22)}
and~\eqref{eq:(2.23)}. Thus, we see that the Feynman rules of the above
$(D+1)$-dimensional field theory basically coincide with the perturbative
expansion of the flowed system in~Sect.~\ref{sec:2.3}.

A possible subtlety arises, however, for a Feynman loop consisting only of the
$BL$~propagators (and similarly for a loop consisting only of the
$\chi\Bar{\lambda}$ and $\lambda\Bar{\chi}$ propagators), because although we
can write such a Feynman diagram in the $(D+1)$-dimensional field theory, we
have no such loop diagram of the flow lines (heat kernels) in the perturbative
expansion of the flowed system, as we have noted at the end
of~Sect.~\ref{sec:2.3}. In reality, this possible discrepancy does not arise
according to the above forward difference prescription~\eqref{eq:(3.22)},
because we have for example $\langle B_\mu^a(t,x)L_\nu^b(s,y)\rangle_0=0$
for~$t\leq s$. This implies that such a loop consisting only of the $BL$
propagators (and a similar loop of the $\chi\Bar{\lambda}$
or~$\lambda\Bar{\chi}$ propagators) identically vanishes; this is the role of
the forward difference prescription~\eqref{eq:(3.22)}.

A similar remark applies to a loop consisting of the $d\Bar{d}$~propagators.
The interaction vertex in~$S_{d\Bar{d}}$ in~Eq.~\eqref{eq:(3.3)} does not
contribute to correlation function without containing $d$ and~$\Bar{d}$,
because any loop consisting of the $d\Bar{d}$~propagator vanishes because of
the property~$\langle d(t,x)\Bar{d}(s,y)\rangle_0=0$ for~$t\leq s$. Note that
we are interested only in correlation functions without containing $d$
and~$\Bar{d}$; these fields are auxiliary tools simply to derive the WT
relation associated with the BRS symmetry.

This completes our argument for the perturbative equivalence of the
$(D+1)$-dimensional field theory and the system defined by the flow equations.
We can now study the structure of the renormalization in the flowed system, by
employing the $(D+1)$-dimensional local field theory.

\section{Proof of the renormalizability}
\label{sec:4}
In this section, we give a proof of the renormalizability of the flowed system
on the basis of the $(D+1)$-dimensional local field theory introduced in
the previous section. To avoid our basic reasoning from being made obscure by
too complicated mathematical expressions, we first extensively study the case
of the pure Yang--Mills theory. Then, in~Sect.~\ref{sec:4.5}, we explain how
the proof is modified by the inclusion of fermion fields.

\subsection{Ward--Takahashi relation or the Zinn-Justin equation}
\label{sec:4.1}
In this subsection, we derive the WT relation associated with the BRS symmetry
(the Zinn-Justin equation) of the $(D+1)$-dimensional field theory. The
argument is standard, except a caution associated with non-trivial relations
among field variables, Eqs.~\eqref{eq:(3.4)} and~\eqref{eq:(3.5)}.

First, we introduce the external sources for each field by
\begin{align}
   S_J
   &=2\int d^Dx\,
   \tr\left[
   J_\mu^A(x)A_\mu(x)+J^c(x)c(x)+J^{\Bar{c}}(x)\Bar{c}(x)+J^B(x)B(x)\right]
\notag\\
   &\qquad{}
   +2\fint_0^\infty dt\,\int d^Dx\,
   \tr\left[
   J_\mu^B(t,x)B_\mu(t,x)+J^d(t,x)d(t,x)\right]
\notag\\
   &\qquad\qquad{}
   +2\int_0^\infty dt\,\int d^Dx\,
   \tr\left[
   J_\mu^L(t,x)L_\mu(t,x)+J^{\Bar{d}}(t,x)\Bar{d}(t,x)\right].
\label{eq:(4.1)}
\end{align}
In writing this, we have noted that the field $B_\mu(t=0,x)$ is not an
independent dynamical variable but rather $B_\mu(t=0,x)=A_\mu(x)$. Thus, we
should not include the external source~$J_\mu^B(t,x)$ for~$B_\mu(t,x)$ at~$t=0$,
$J_\mu^B(t=0,x)$, because $J_\mu^B(t=0,x)$ cannot be distinguished
from~$J_\mu^A(x)$. In order to take this point into account, we have introduced
a flow-time integration with the $t=0$ time-slice being excluded:
\begin{equation}
   \fint_0^\infty dt\equiv\epsilon\sum_{t=\epsilon}^\infty.
\end{equation}
This remark is applied also to the external source~$J^d(t,x)$ for~$d(t,x)$,
because $d(t=0,x)=c(x)$ is not an independent variable.

We next introduce another source term for the BRS transformations of each
field:
\begin{align}
   S_K
   &=2\int d^Dx\,
   \tr\left[
   K_\mu^A(x)D_\mu c(x)-K^c(x)c(x)^2\right]
\notag
\\
   &\qquad{}
   +2\fint_0^\infty dt\,\int d^Dx\,
   \tr\left[
   K_\mu^B(t,x)D_\mu d(t,x)-K^d(t,x)d(t,x)^2\right]
\notag
\\
   &\qquad\qquad{}
   +2\int_0^\infty dt\,\int d^Dx\,
   \tr\Bigl\{
   K_\mu^L(t,x)[L_\mu(t,x),d(t,x)]
\notag
\\
   &\qquad\qquad\qquad\qquad\qquad\qquad\qquad{}
   +K^{\Bar{d}}(t,x)\left[D_\mu L_\mu(t,x)
   -\{d(t,x),\Bar{d}(t,x)\}\right]\Bigr\}.
\label{eq:(4.3)}
\end{align}
The symbol~$\fint_0^\infty dt$ is used for the same reason as
for~Eq.~\eqref{eq:(4.1)}. Note that because of the nilpotency of the BRS
transformation~$\delta^2=0$, $\delta S_K=0$.

Then, as usual, we consider the BRS transformation~$\delta$ of integration
variables in the functional integral
\begin{equation}
   e^{-W[J,K]}\equiv\int d\mu\,e^{-S_{\text{tot}}-S_J-S_K},
\label{eq:(4.4)}
\end{equation}
where $d\mu$ is the functional integration measure. Then, since
$S_{\text{tot}}$~\eqref{eq:(3.1)} and~$S_K$~\eqref{eq:(4.3)} are invariant
under~$\delta$ while $S_J$~\eqref{eq:(4.1)} is not, we have the following
identity among correlation functions:
\begin{align}
   &\left\langle
   -2\int d^Dx\,
   \tr\left[
   J_\mu^A(x)D_\mu c(x)+J^c(x)c(x)^2-J^{\Bar{c}}(x)B(x)\right]
   \right\rangle
\notag
\\
   &\qquad{}
   +\left\langle-2\fint_0^\infty dt\,\int d^Dx\,
   \tr\left[
   J_\mu^B(t,x)D_\mu d(t,x)+J^d(t,x)d(t,x)^2\right]\right\rangle
\notag
\\
   &\qquad\qquad{}
   +\biggl\langle-2\int_0^\infty dt\,\int d^Dx\,
   \tr\Bigl\{
   J_\mu^L(t,x)[L_\mu(t,x),d(t,x)]
\notag\\
   &\qquad\qquad\qquad\qquad\qquad\qquad\qquad\qquad{}
   -J^{\Bar{d}}\left[D_\mu L_\mu(t,x)
   -\{d(t,x),\Bar{d}(t,x)\}\right]\Bigr\}
   \biggr\rangle=0.
\label{eq:(4.5)}
\end{align}
In reality, this identity is broken by~$O(\epsilon)$ terms under the discrete
flow-time prescription that we are assuming at present. To explicitly see those
$O(\epsilon)$ terms safely vanish for the $\epsilon\to0$ limit requires a
little work and Appendix~\ref{app:B} is devoted to this task. In what follows,
we neglect those $O(\epsilon)$ terms by assuming that the limit $\epsilon\to0$
is taken in an appropriate stage.\footnote{Another possible approach is to
consider a lattice regularization of the $(D+1)$-dimensional field
theory~\cite{Luscher:2013cpa} in which the BRS symmetry is manifest and then
study the perturbative renormalizability by using this regularization.}

We then introduce the effective action~$\mathit{\Gamma}$, the generating
functional of one-particle irreducible (1PI) diagrams in the presence of
$K$~external sources, by the Legendre transformation from external sources to
the expectation values by,
\begin{align}
   &\mathit{\Gamma}[A_\mu,c,\Bar{c},B,\Tilde{K}_\mu^A,K^c;
   B_\mu,d,L_\mu,\Bar{d},K_\mu^B,K^d,K_\mu^L,K^{\Bar{d}}]
\notag\\
   &\equiv W[J,K]
\notag\\
   &\qquad{}
   +2\int d^Dx\,
   \tr\left[
   J_\mu^A(x)A_\mu(x)+J^c(x)c(x)+J^{\Bar{c}}(x)\Bar{c}(x)+J^B(x)B(x)\right]
\notag\\
   &\qquad\qquad{}
   +2\fint_0^\infty dt\,\int d^Dx\,
   \tr\left[
   J_\mu^B(t,x)B_\mu(t,x)+J^d(t,x)d(t,x)\right]
\notag\\
   &\qquad\qquad\qquad{}
   +2\int_0^\infty dt\,\int d^Dx\,
   \tr\left[
   J_\mu^L(t,x)L_\mu(t,x)+J^{\Bar{d}}(t,x)\Bar{d}(t,x)\right],
\label{eq:(4.6)}
\end{align}
where field variables in this expression, such as $A_\mu(x)$, stand for
expectation values rather than dynamical variables. Note that the $K$~external
sources are not Legendre transformed; the dependence on~$K$ is thus taken over
from~$W$ to~$\mathit{\Gamma}$. Then, noting the relation such
as~$\delta\mathit{\Gamma}/\delta A_\mu^a(x)=J_\mu^a(x)$, the WT
relation~\eqref{eq:(4.5)} is expressed in terms of the effective
action~$\mathit{\Gamma}$ as
\begin{align}
   &\int d^Dx\,
   \left[
   \frac{\delta\mathit{\Gamma}}{\delta A_\mu^a(x)}
   \frac{\delta\mathit{\Gamma}}{\delta K_\mu^{Aa}(x)}
   +\frac{\delta\mathit{\Gamma}}{\delta c^a(x)}
   \frac{\delta\mathit{\Gamma}}{\delta K^{ca}(x)}
   -\frac{\delta\mathit{\Gamma}}{\delta\Bar{c}^a(x)}B^a(x)
   \right]
\notag\\
   &\qquad{}
   +\fint_0^\infty dt\,\int d^Dx\,
   \left[
   \frac{\delta\mathit{\Gamma}}{\delta B_\mu^a(t,x)}
   \frac{\delta\mathit{\Gamma}}{\delta K_\mu^{Ba}(t,x)}
   +\frac{\delta\mathit{\Gamma}}{\delta d^a(t,x)}
   \frac{\delta\mathit{\Gamma}}{\delta K^{da}(t,x)}
   \right]
\notag\\
   &\qquad\qquad{}
   +\int_0^\infty dt\,\int d^Dx\,
   \left[
   \frac{\delta\mathit{\Gamma}}{\delta L_\mu^a(t,x)}
   \frac{\delta\mathit{\Gamma}}{\delta K_\mu^{La}(t,x)}
   +\frac{\delta\mathit{\Gamma}}{\delta\Bar{d}^a(t,x)}
   \frac{\delta\mathit{\Gamma}}{\delta K^{\Bar{d}a}(t,x)}
   \right]
   =0.
\label{eq:(4.7)}
\end{align}

Next, we note that the equations of motion (the Schwinger--Dyson equations),
\begin{align}
   \left\langle
   \frac{1}{g_0^2}\left[\partial_\mu A_\mu^a(x)-\frac{1}{\lambda_0}B^a(x)\right]
   \right\rangle
   -J^{Ba}(x)&=0,
\label{eq:(4.8)}
\\
   \left\langle
   -\frac{1}{g_0^2}\partial_\mu D_\mu c^a(x)\right\rangle
   +J^{\Bar{c}a}(x)&=0,
\label{eq:(4.9)}
\end{align}
are expressed in terms of the effective action as
\begin{align}
   &\frac{\delta\mathit{\Gamma}}{\delta B^a(x)}
   =\frac{1}{g_0^2}\left[\partial_\mu A_\mu^a(x)-\frac{1}{\lambda_0}B^a(x)
   \right],
\label{eq:(4.10)}
\\
   &\frac{\delta\mathit{\Gamma}}{\delta\Bar{c}^a(x)}
   -\frac{1}{g_0^2}\partial_\mu\frac{\delta\mathit{\Gamma}}{\delta K_\mu^{Aa}(x)}
   =0.
\label{eq:(4.11)}
\end{align}
These relations show that, by defining
\begin{equation}
   \Tilde{K}_\mu^{Aa}\equiv K_\mu^{Aa}-\frac{1}{g_0^2}\partial_\mu\Bar{c}^a
\label{eq:(4.12)}
\end{equation}
and
\begin{align}
   &\Tilde{\mathit{\Gamma}}[A_\mu,c,\Tilde{K}_\mu^A,K^c;
   B_\mu,d,L_\mu,\Bar{d},K_\mu^B,K^d,K_\mu^L,K^{\Bar{d}}]
\notag\\
   &\equiv\mathit{\Gamma}
   +\frac{2}{g_0^2}\int d^Dx\,
   \tr\left\{
   B(x)\left[\partial_\mu A_\mu(x)-\frac{1}{2\lambda_0}B(x)
   \right]\right\},
\label{eq:(4.13)}
\end{align}
the dependence on~$B$ and~$\Bar{c}$ are completely eliminated;
$\Tilde{\mathit{\Gamma}}$ does not depend on neither~$B$ nor~$\Bar{c}$. The
reduced effective action~$\Tilde{\mathit{\Gamma}}$ then obeys
\begin{align}
   &\int d^Dx\,
   \left[
   \frac{\delta\Tilde{\mathit{\Gamma}}}{\delta A_\mu^a(x)}
   \frac{\delta\Tilde{\mathit{\Gamma}}}{\delta \Tilde{K}_\mu^{Aa}(x)}
   +\frac{\delta\Tilde{\mathit{\Gamma}}}{\delta c^a(x)}
   \frac{\delta\Tilde{\mathit{\Gamma}}}{\delta K^{ca}(x)}
   \right]
\notag\\
   &\qquad{}
   +\fint_0^\infty dt\,\int d^Dx\,
   \left[
   \frac{\delta\Tilde{\mathit{\Gamma}}}{\delta B_\mu^a(t,x)}
   \frac{\delta\Tilde{\mathit{\Gamma}}}{\delta K_\mu^{Ba}(t,x)}
   +\frac{\delta\Tilde{\mathit{\Gamma}}}{\delta d^a(t,x)}
   \frac{\delta\Tilde{\mathit{\Gamma}}}{\delta K^{da}(t,x)}
   \right]
\notag\\
   &\qquad\qquad{}
   +\int_0^\infty dt\,\int d^Dx\,
   \left[
   \frac{\delta\Tilde{\mathit{\Gamma}}}{\delta L_\mu^a(t,x)}
   \frac{\delta\Tilde{\mathit{\Gamma}}}{\delta K_\mu^{La}(t,x)}
   +\frac{\delta\Tilde{\mathit{\Gamma}}}{\delta\Bar{d}^a(t,x)}
   \frac{\delta\Tilde{\mathit{\Gamma}}}{\delta K^{\Bar{d}a}(t,x)}
   \right]
   =0.
\label{eq:(4.14)}
\end{align}
This is the WT relation following from the BRS invariance of the
$(D+1)$-dimensional field theory.

\subsection{Statement of the renormalizability}
\label{sec:4.2}
Now, our statement of renormalizability is that the generating functional of
the 1PI correlation functions~$\mathit{\Gamma}$ can be made finite in terms of
renormalized quantities, by choosing three renormalization
constants\footnote{When fermion fields are included, we need additional
renormalization constants, $Z_\psi$, $Z_m$, and~$Z_\chi$;
see~Sect.~\ref{sec:4.5}.} $Z$, $Z_3$, and~$\Tilde{Z}_3$ in
\begin{align}
   g_0^2&=\mu^{2\varepsilon}g^2Z,&&
\label{eq:(4.15)}
\\
   A_\mu^a&=Z^{1/2}Z_3^{1/2}(A_R)_\mu^a,&
   K_\mu^{Aa}&=Z^{-1/2}Z_3^{-1/2}(K_R^A)_\mu^a,
\label{eq:(4.16)}
\\
   c^a&=\Tilde{Z}_3Z^{1/2}Z_3^{1/2}c_R^a,
   &K^{ca}&=\Tilde{Z}_3^{-1}Z^{-1/2}Z_3^{-1/2}K_R^{ca},
\label{eq:(4.17)}
\end{align}
and
\begin{align}
   \lambda_0&=\lambda Z_3^{-1},&&
\label{eq:(4.18)}
\\
   B^a&=Z^{1/2}Z_3^{-1/2}B_R^a,&\Bar{c}^a&=Z^{1/2}Z_3^{-1/2}\Bar{c}_R^a,
\label{eq:(4.19)}
\end{align}
appropriately order by order in the loop expansion of perturbation theory.
Note that the above statement in particular claims that the flowed gauge
field~$B_\mu(t,x)$ \emph{does not require any wave function
renormalization}.\footnote{It also claims that the parameter~$\alpha_0$ is not
renormalized.} From~Eq.~\eqref{eq:(4.13)}, it suffices to show the finiteness
of the reduced effective action~$\Tilde{\mathit{\Gamma}}$, because
\begin{equation}
   -\frac{2}{g_0^2}
   \tr\left[
   B\left(\partial_\mu A_\mu-\frac{1}{2\lambda_0}B
   \right)\right]
   =
   -\frac{2}{\mu^{2\varepsilon}g^2}
   \tr\left\{
   B_R\left[\partial_\mu (A_R)_\mu-\frac{1}{2\lambda}B_R
   \right]\right\},
\label{eq:(4.20)}
\end{equation}
is finite with the above definitions. For $\Tilde{\mathit{\Gamma}}$, defining
\begin{equation}
   \Tilde{K}_\mu^{Aa}=K_\mu^{Aa}-\frac{1}{g_0^2}\partial_\mu\Bar{c}^a
   \equiv Z^{-1/2}Z_3^{-1/2}(\Tilde{K}_R^A)_\mu^a,
\label{eq:(4.21)}
\end{equation}
the WT relation~\eqref{eq:(4.14)} in terms of the renormalized quantities
yields,
\begin{align}
   &\int d^Dx\,
   \left[
   \frac{\delta\Tilde{\mathit{\Gamma}}}{\delta(A_R)_\mu^a(x)}
   \frac{\delta\Tilde{\mathit{\Gamma}}}{\delta(\Tilde{K}_R^A)_\mu^a(x)}
   +\frac{\delta\Tilde{\mathit{\Gamma}}}{\delta c_R^a(x)}
   \frac{\delta\Tilde{\mathit{\Gamma}}}{\delta K_R^{ca}(x)}
   \right]
\notag\\
   &\qquad{}
   +\fint_0^\infty dt\,\int d^Dx\,
   \left[
   \frac{\delta\Tilde{\mathit{\Gamma}}}{\delta B_\mu^a(t,x)}
   \frac{\delta\Tilde{\mathit{\Gamma}}}{\delta K_\mu^{Ba}(t,x)}
   +\frac{\delta\Tilde{\mathit{\Gamma}}}{\delta d^a(t,x)}
   \frac{\delta\Tilde{\mathit{\Gamma}}}{\delta K^{da}(t,x)}
   \right]
\notag\\
   &\qquad\qquad{}
   +\int_0^\infty dt\,\int d^Dx\,
   \left[
   \frac{\delta\Tilde{\mathit{\Gamma}}}{\delta L_\mu^a(t,x)}
   \frac{\delta\Tilde{\mathit{\Gamma}}}{\delta K_\mu^{La}(t,x)}
   +\frac{\delta\Tilde{\mathit{\Gamma}}}{\delta\Bar{d}^a(t,x)}
   \frac{\delta\Tilde{\mathit{\Gamma}}}{\delta K^{\Bar{d}a}(t,x)}
   \right]
   =0.
\label{eq:(4.22)}
\end{align}

Let us now consider the loop expansion of~$\Tilde{\mathit{\Gamma}}$ in the
renormalized perturbation theory,
\begin{equation}
   \Tilde{\mathit{\Gamma}}=\sum_{\ell=0}^\infty\Tilde{\mathit{\Gamma}}^{(\ell)},
\label{eq:(4.23)}
\end{equation}
where $\ell$ denotes the order of the loop expansion; the tree-level effective
action~$\Tilde{\mathit{\Gamma}}^{(0)}$ is given by the tree-level action in
renormalized quantities:
\begin{equation}
   \Tilde{\mathit{\Gamma}}^{(0)}
   =\left.\left(S+S_{\text{fl}}+S_{d\Bar{d}}+S_K\right)\right|
   _{Z=Z_3=\Tilde{Z_3}=1,K_\mu^A\to(\Tilde{K}_R^A)_\mu}.
\label{eq:(4.24)}
\end{equation}
Corresponding to the tree-level action~\eqref{eq:(4.24)}, the counterterm in
the renormalized perturbation theory is given by
\begin{equation}
   \Delta S
   =S+S_{\text{fl}}+S_{d\Bar{d}}+S_K
   -\left.\left(S+S_{\text{fl}}+S_{d\Bar{d}}+S_K\right)\right|
   _{Z=Z_3=\Tilde{Z_3}=1,K_\mu^A\to(\Tilde{K}_R^A)_\mu}.
\label{eq:(4.25)}
\end{equation}
Note that this counterterm contains in particular the ``boundary counterterm''
$\Delta S_{\text{bc}}$~\cite{Luscher:2011bx},
\begin{align}
   \Delta S_{\text{bc}}
   &=2\int d^Dx\,\tr\left[L_\mu(0,x)(Z^{1/2}Z_3^{1/2}-1)
   (A_R)_\mu(x)\right]
\notag\\
   &\qquad{}
   +2\int d^Dx\,\tr\left[\Bar{d}(0,x)(\Tilde{Z}_3Z^{1/2}Z_3^{1/2}-1)
   c_R(x)\right],
\label{eq:(4.26)}
\end{align}
because in~Eqs.~\eqref{eq:(3.2)} and~\eqref{eq:(3.3)}, the $t$-derivatives
at~$t=0$ are defined by Eqs.~\eqref{eq:(3.23)} and~\eqref{eq:(3.25)} and
$A_\mu(x)$ and~$c(x)$ are renormalized as~$Z^{1/2}Z_3^{1/2}(A_R)_\mu(x)$
and~$\Tilde{Z}_3Z^{1/2}Z_3^{1/2}c_R(x)$, while $B_\mu(t=\epsilon,x)$
and~$d(t=\epsilon,x)$ are not renormalized. We also note that the tree-level
action~\eqref{eq:(4.24)} yields the tree-level propagators
\begin{equation}
   \left\langle (A_R)_\mu^a(x)(A_R)_\nu^b(y)\right\rangle_0
   =\mu^{2\varepsilon}g^2\delta^{ab}
   \int_p\frac{e^{ip(x-y)}}{(p^2)^2}
   \left(\delta_{\mu\nu}p^2-p_\mu p_\nu
   +\frac{1}{\lambda}p_\mu p_\nu\right),
\label{eq:(4.27)}
\end{equation}
and (in the $\epsilon\to0$ limit)
\begin{equation}
   \left\langle B_\mu^a(t,x)B_\nu^b(s,y)\right\rangle_0
   =\mu^{2\varepsilon}g^2\delta^{ab}
   \int_p\frac{e^{ip(x-y)}}{(p^2)^2}
   \left[(\delta_{\mu\nu}p^2-p_\mu p_\nu)e^{-(t+s)p^2}
   +\frac{1}{\lambda}p_\mu p_\nu e^{-\alpha_0(t+s)p^2}\right].
\label{eq:(4.28)}
\end{equation}
In the renormalized perturbation theory, these propagators are used.

Our proof for the renormalizability proceeds by induction: Suppose that, to the
$\ell$-th order of the loop expansion, $Z$, $Z_3$, and~$\Tilde{Z}_3$ can be
chosen so that $\Tilde{\mathit{\Gamma}}^{(\ell)}$ is finite. Then, considering
the divergent part of the WT relation~\eqref{eq:(4.22)} in the $(\ell+1)$-th
loop order, we have
\begin{equation}
   \Tilde{\mathit{\Gamma}}^{(0)}\ast\Tilde{\mathit{\Gamma}}^{(\ell+1)\text{div}}
   =0,
\label{eq:(4.29)}
\end{equation}
where $\Tilde{\mathit{\Gamma}}^{(\ell+1)\text{div}}$ denotes the divergent part
of~$\Tilde{\mathit{\Gamma}}^{(\ell+1)}$ and
\begin{align}
   \Tilde{\mathit{\Gamma}}^{(0)}\ast&\equiv
   -\int d^Dx\,
   \Biggl[
   \frac{\delta\Tilde{\mathit{\Gamma}}^{(0)}}{\delta(A_R)_\mu^a(x)}
   \frac{\delta}{\delta(\Tilde{K}_R^A)_\mu^a(x)}
   +\frac{\delta\Tilde{\mathit{\Gamma}}^{(0)}}{\delta(\Tilde{K}_R^A)_\mu^a(x)}
   \frac{\delta}{\delta(A_R)_\mu^a(x)}
\notag\\
   &\qquad\qquad\qquad{}
   +\frac{\delta\Tilde{\mathit{\Gamma}}^{(0)}}{\delta c_R^a(x)}
   \frac{\delta}{\delta K_R^{ca}(x)}
   +\frac{\delta\Tilde{\mathit{\Gamma}}^{(0)}}{\delta K_R^{ca}(x)}
   \frac{\delta}{\delta c_R^a(x)}
   \Biggr]
\notag\\
   &\qquad{}
   -\epsilon\int d^Dx\,
   \Biggl[
   \frac{\delta\Tilde{\mathit{\Gamma}}^{(0)}}{\delta L_\mu^a(0,x)}
   \frac{\delta}{\delta K_\mu^{La}(0,x)}
   +\frac{\delta\Tilde{\mathit{\Gamma}}^{(0)}}{\delta K_\mu^{La}(0,x)}
   \frac{\delta}{\delta L_\mu^a(0,x)}
\notag\\
   &\qquad\qquad\qquad\qquad{}
   +\frac{\delta\Tilde{\mathit{\Gamma}}^{(0)}}{\delta\Bar{d}^a(0,x)}
   \frac{\delta}{\delta K^{\Bar{d}a}(0,x)}
   +\frac{\delta\Tilde{\mathit{\Gamma}}^{(0)}}{\delta K^{\Bar{d}a}(0,x)}
   \frac{\delta}{\delta\Bar{d}^a(0,x)}
   \Biggr]
\notag\\
   &\qquad\qquad\qquad\qquad\qquad{}
   +(\text{derivatives w.r.t.\ field variables at $t\neq0$}).
\label{eq:(4.30)}
\end{align}
In this expression, we have explicitly used the discrete time
prescription~\eqref{eq:(3.21)}.\footnote{A $(D+1)$-dimensional functional
derivative, such as $\delta/\delta K_\mu^{La}(0,x)$, should be read
as~$(1/\epsilon)\delta/\delta K_\mu^{La}(0,x)$, when acting on a
$D$-dimensional integral such as $\int d^Dx\,K_\mu^{La}(0,x)f(x)$. We also
discard terms being higher order in~$\epsilon$ in following calculations.} One
can confirm that because of the BRS invariance of the
action~$S+S_{\text{fl}}+S_{d\Bar{d}}+S_K$, $\Tilde{\mathit{\Gamma}}^{(0)}\ast$ is
nilpotent:
\begin{equation}
    \Tilde{\mathit{\Gamma}}^{(0)}\ast\Tilde{\mathit{\Gamma}}^{(0)}\ast=0.
\label{eq:(4.31)}
\end{equation}
Eqs.~\eqref{eq:(4.29)} and~\eqref{eq:(4.31)} are basic relations for the proof.
Next, we have to clarify the most general form of the divergent
part~$\Tilde{\mathit{\Gamma}}^{(\ell+1)\text{div}}$.

\subsection{General form of the divergent part}
\label{sec:4.3}
First of all, we note that there is no ``bulk'' divergence that is given by a
$(D+1)$-dimensional integral $\int_0^\infty dt\int d^Dx$ of a local polynomial
of fields. This follows from the fact that a Feynman loop that consists only of
the flow lines (the heat kernels) identically vanishes as we have observed.
Thus the integrand of the loop integral for a loop diagram whose vertices
reside in the ``bulk'' of the $(D+1)$-dimensional spacetime, $t>0$, contains at
least one propagator which provides the Gaussian damping
factor~$\sim e^{-tp^2}$. Then the loop-momentum integral is absolutely
convergent; there is no ``bulk'' divergence.

Thus $\Tilde{\mathit{\Gamma}}^{(\ell+1)\text{div}}$ consists of the
$D$-dimensional integral~$\int d^Dx$ of a local polynomial of fields, the
``boundary'' divergence. The mass dimension of the local polynomial is less or
equal to~$4$ for~$D=4$ and the ghost number is~$0$.

Let us next show that
\begin{equation}
   -2\int d^Dx\,\tr\left[\partial_tB_\mu(0,x)(A_R)_\mu(x)\right],\qquad
   -2\int d^Dx\,\tr\left[\partial_td(0,x)\Bar{c}_R(x)\right],
\label{eq:(4.32)}
\end{equation}
cannot appear in~$\Tilde{\mathit{\Gamma}}^{(\ell+1)\text{div}}$, although the mass
dimension and the ghost number perfectly match. Note that two combinations
in~Eq.~\eqref{eq:(4.32)} contain, according to our prescription,
$B_\mu(t=\epsilon,x)$ and~$d(t=\epsilon,x)$, respectively. That the terms
in~Eq.~\eqref{eq:(4.32)} cannot appear in the divergent part follows again from
the fact that a Feynman loop that consists only of the flow lines, the $BL$ and
$d\Bar{d}$ propagators, identically vanishes. Because of this fact and from the
structure of the interaction vertices containing $B_\mu$ or~$d$
in~$S_{\text{tot}}+S_K$, we see that any divergent term that contains $B_\mu$
or~$d$ must involve $L_\mu$, $\Bar{d}$, or the external source~$K$. The
combinations in~Eq.~\eqref{eq:(4.32)} do not satisfy this
criterion.\footnote{One has to note this property of the divergence part also
in the proof of~Ref.~\cite{Luscher:2011bx}. We would like to thank Martin
L\"uscher for explanation on this point.}

Next, consider terms being proportional to the flow-time derivative at~$t=0$,
$\partial_tB_\mu^a(0,x)$ or~$\partial_td^a(0,x)$ in the WT
relation~$\Tilde{\mathit{\Gamma}}^{(0)}\ast\Tilde{\mathit{\Gamma}}^{(\ell+1)\text{div}}=0$. In~$\Tilde{\mathit{\Gamma}}^{(0)}*$ in~Eq.~\eqref{eq:(4.30)}, only the
coefficient functions,
\begin{equation}
   \frac{\delta\Tilde{\mathit{\Gamma}}^{(0)}}{\delta L_\mu^a(0,x)}
   =\partial_tB_\mu^a(0,x)+\dotsb,\qquad
   \frac{\delta\Tilde{\mathit{\Gamma}}^{(0)}}{\delta\Bar{d}^a(0,x)}
   =\partial_td^a(0,x)+\dotsb,
\label{eq:(4.33)}
\end{equation}
contain the flow-time derivative at~$t=0$. Since
$\Tilde{\mathit{\Gamma}}^{(0)}*$ has the structure
\begin{equation}
   \Tilde{\mathit{\Gamma}}^{(0)}*
   \sim\frac{\delta\Tilde{\mathit{\Gamma}}^{(0)}}{\delta L_\mu^a(0,x)}
   \frac{\delta}{\delta K_\mu^{La}(0,x)}
   +\frac{\delta\Tilde{\mathit{\Gamma}}^{(0)}}{\delta\Bar{d}^a(0,x)}
   \frac{\delta}{\delta K^{\Bar{d}a}(0,x)}+\dotsb,
\label{eq:(4.34)}
\end{equation}
taking into account the fact that $\Tilde{\mathit{\Gamma}}^{(\ell+1)\text{div}}$
contains neither $\partial_tB_\mu^a(0,x)$ nor~$\partial_td^a(0,x)$ as we have
observed above, we infer that $\Tilde{\mathit{\Gamma}}^{(\ell+1)\text{div}}$
cannot contain external fields, $K_\mu^L(0,x)$ and~$K^{\Bar{d}}(0,x)$.

Finally, the mass dimension for~$D=4$ and the ghost number of~$K_\mu^B$ ($K^d$)
are $4$ and~$-1$ ($4$ and~$-2$), respectively. For this reason, these sources 
cannot appear in~$\Tilde{\mathit{\Gamma}}^{(\ell+1)\text{div}}$.

From these considerations, we find that the most general form of the divergent
part is given by
\begin{equation}
   \Tilde{\mathit{\Gamma}}^{(\ell+1)\text{div}}
   =\int d^Dx\,\mathcal{L}
   \left((A_R)_\mu,c_R,(\Tilde{K}_R^A)_\mu,K_R^c;L_\mu(t=0),\Bar{d}(t=0)\right),
\label{eq:(4.35)}
\end{equation}
where $\mathcal{L}$ is a local polynomial of fields; its mass dimension is at
most~$4$ for~$D=4$ and the ghost number is~$0$.

\subsection{Final steps}
\label{sec:4.4}
Our strategy to analyze the WT relation~\eqref{eq:(4.29)} is to decompose it
into a $D$-dimensional part that is well-understood in the renormalization of
the Yang--Mills theory and an exotic part that originates from the
$(D+1)$-dimensional fields.

Thus, in~Eq.~\eqref{eq:(4.35)}, we set
$\Tilde{\mathit{\Gamma}}^{(\ell+1)\text{div}}=\Tilde{\mathit{\Gamma}}_D^{(\ell+1)\text{div}}+\Tilde{\mathit{\Gamma}}_{D+1}^{(\ell+1)\text{div}}$, where
\begin{align}
   \Tilde{\mathit{\Gamma}}_D^{(\ell+1)\text{div}}
   &\equiv
   \left.\Tilde{\mathit{\Gamma}}^{(\ell+1)\text{div}}\right|_{L_\mu=\Bar{d}=0},
\label{eq:(4.36)}
\\
   \Tilde{\mathit{\Gamma}}_{D+1}^{(\ell+1)\text{div}}
   &\equiv\Tilde{\mathit{\Gamma}}^{(\ell+1)\text{div}}
   -\Tilde{\mathit{\Gamma}}_D^{(\ell+1)\text{div}}.
\label{eq:(4.37)}
\end{align}
Note that $\Tilde{\mathit{\Gamma}}_{D+1}^{(\ell+1)\text{div}}$ contains at least
one $L_\mu$ or one $\Bar{d}$.

We also decompose $\Tilde{\mathit{\Gamma}}^{(0)}\ast$ into
$\Tilde{\mathit{\Gamma}}^{(0)}\ast
=\Tilde{\mathit{\Gamma}}_D^{(0)}\ast+\Tilde{\mathit{\Gamma}}_{D+1}^{(0)}\ast$,
where $\Tilde{\mathit{\Gamma}}_D^{(0)}\ast$ is the ``BRS operator'' in the
original Yang--Mills theory
\begin{align}
   \Tilde{\mathit{\Gamma}}_D^{(0)}\ast&\equiv
   -\int d^Dx\,
   \Biggl[
   \frac{\delta\left.
   \Tilde{\mathit{\Gamma}}^{(0)}
   \right|_{L_\mu=\Bar{d}=0}}{\delta(A_R)_\mu^a(x)}
   \frac{\delta}{\delta(\Tilde{K}_R^A)_\mu^a(x)}
   +\frac{\delta\left.
   \Tilde{\mathit{\Gamma}}^{(0)}
   \right|_{L_\mu=\Bar{d}=0}}{\delta(\Tilde{K}_R^A)_\mu^a(x)}
   \frac{\delta}{\delta(A_R)_\mu^a(x)}
\notag\\
   &\qquad\qquad\qquad{}
   +\frac{\delta\left.
   \Tilde{\mathit{\Gamma}}^{(0)}
   \right|_{L_\mu=\Bar{d}=0}}{\delta c_R^a(x)}
   \frac{\delta}{\delta K_R^{ca}(x)}
   +\frac{\delta\left.
   \Tilde{\mathit{\Gamma}}^{(0)}
   \right|_{L_\mu=\Bar{d}=0}}{\delta K_R^{ca}(x)}
   \frac{\delta}{\delta c_R^a(x)}
   \Biggr]
\label{eq:(4.38)}
\end{align}
and $\Tilde{\mathit{\Gamma}}_{D+1}^{(0)}\ast$ is the remaining
\begin{equation}
   \Tilde{\mathit{\Gamma}}_{D+1}^{(0)}\ast
   \equiv\Tilde{\mathit{\Gamma}}^{(0)}\ast
   -\Tilde{\mathit{\Gamma}}_D^{(0)}\ast.
\label{eq:(4.39)}
\end{equation}

Under the above decompositions, we see that the WT relation~\eqref{eq:(4.29)}
is decomposed as
\begin{align}
   &\Tilde{\mathit{\Gamma}}_D^{(0)}\ast
   \Tilde{\mathit{\Gamma}}_D^{(\ell+1)\text{div}}=0,
\label{eq:(4.40)}
\\
   &\Tilde{\mathit{\Gamma}}_D^{(0)}\ast
   \Tilde{\mathit{\Gamma}}_{D+1}^{(\ell+1)\text{div}}
   +
   \Tilde{\mathit{\Gamma}}_{D+1}^{(0)}\ast
   \Tilde{\mathit{\Gamma}}_D^{(\ell+1)\text{div}}
   +
   \Tilde{\mathit{\Gamma}}_{D+1}^{(0)}\ast
   \Tilde{\mathit{\Gamma}}_{D+1}^{(\ell+1)\text{div}}=0.
\label{eq:(4.41)}
\end{align}
To conclude this, we have to note that
$\Tilde{\mathit{\Gamma}}_{D+1}^{(\ell+1)\text{div}}$ is independent of~$K_\mu^L$
and~$K^{\Bar{d}}$ and thus
$\delta\Tilde{\mathit{\Gamma}}^{(0)}/\delta L_\mu^a(0,x)$ and
$\delta\Tilde{\mathit{\Gamma}}^{(0)}/\delta\Bar{d}^a(0,x)$
in~$\Tilde{\mathit{\Gamma}}_{D+1}^{(0)}\ast$, which might produce terms
independent of~$L_\mu$ and~$\Bar{d}$, do not contribute to the WT relation.

Now, the general solution to Eq.~\eqref{eq:(4.40)} is well-known; it is given
by the sum of a gauge invariant part and the ``BRS trivial
part''~\cite{Joglekar:1975nu}:
\begin{align}
   \Tilde{\mathit{\Gamma}}_D^{(\ell+1)\text{div}}
   &=-\frac{1}{2\mu^{2\varepsilon}g^2}
   \int d^Dx\,\tr\left[x_1[F_R]_{\mu\nu}(x)[F_R]_{\mu\nu}(x)\right]
\notag\\
   &\qquad{}
   -2\Tilde{\mathit{\Gamma}}_D^{(0)}\ast
   \int d^Dx\,\tr\left[y_1(\Tilde{K}_R^A)_\mu(x)(A_R)_\mu(x)
   +y_2K_R^c(x)c_R(x)\right],
\label{eq:(4.42)}
\end{align}
where $x_1$, $y_1$, and~$y_2$ are (divergent) constants, and
\begin{equation}
   [F_R]_{\mu\nu}(x)\equiv
   \partial_\mu(A_R)_\nu(x)-\partial_\nu(A_R)_\mu(x)
   +[(A_R)_\mu,(A_R)_\nu](x).
\label{eq:(4.43)}
\end{equation}
Substituting this in~Eq.~\eqref{eq:(4.37)} and using Eq.~\eqref{eq:(4.39)}, we
have
\begin{align}
   \Tilde{\mathit{\Gamma}}^{(\ell+1)\text{div}}
   &=-\frac{1}{2\mu^{2\varepsilon}g^2}
   \int d^Dx\,\tr\left[x_1[F_R]_{\mu\nu}(x)[F_R]_{\mu\nu}(x)\right]
\notag\\
   &\qquad{}
   -2\Tilde{\mathit{\Gamma}}^{(0)}\ast
   \int d^Dx\,\tr\left[y_1(\Tilde{K}_R^A)_\mu(x)(A_R)_\mu(x)
   +y_2K_R^c(x)c_R(x)\right]
\notag\\
   &\qquad\qquad{}
   +\Tilde{\Tilde{\mathit{\Gamma}}}_{D+1}^{(\ell+1)\text{div}},
\label{eq:(4.44)}
\end{align}
where
\begin{equation}
   \Tilde{\Tilde{\mathit{\Gamma}}}_{D+1}^{(\ell+1)\text{div}}
   \equiv
   \Tilde{\mathit{\Gamma}}_{D+1}^{(\ell+1)\text{div}}
   +2\int d^Dx\,\tr\left[y_1L_\mu(0,x)(A_R)_\mu(x)
   -y_2\Bar{d}(0,x)c_R(x)\right].
\label{eq:(4.45)}
\end{equation}
This is the general form of $\Tilde{\mathit{\Gamma}}^{(\ell+1)\text{div}}$,
that is consistent with the WT relation.

$\Tilde{\Tilde{\mathit{\Gamma}}}_{D+1}^{(\ell+1)\text{div}}$ itself is a boundary
integral of a local polynomial of fields, that must contain at least one
$L_\mu$ or one $\Bar{d}$. The general form of such term is
\begin{equation}
   \Tilde{\Tilde{\mathit{\Gamma}}}_{D+1}^{(\ell+1)\text{div}}
   =-2\int d^Dx\,\tr\left[z_1L_\mu(0,x)(A_R)_\mu(x)
   +z_2\Bar{d}(0,x)c_R(x)\right].
\label{eq:(4.46)}
\end{equation}
Also, because of the nilpotency~\eqref{eq:(4.31)}, the WT
relation~\eqref{eq:(4.29)} for~Eq.~\eqref{eq:(4.44)} requires (the gauge
invariant part in the first line of~Eq.~\eqref{eq:(4.44)} is also invariant
under~$\Tilde{\mathit{\Gamma}}^{(0)}\ast$)
\begin{align}
   &\Tilde{\mathit{\Gamma}}^{(0)}\ast
   \Tilde{\Tilde{\mathit{\Gamma}}}_{D+1}^{(\ell+1)\text{div}}
\notag\\
   &=-2\int d^Dx\,
   \tr\bigl[
   (z_1-z_2)L_\mu(0,x)\partial_\mu c_R(x)
\notag\\
   &\qquad\qquad\qquad\qquad{}
   -z_2L_\mu(0,x)[(A_R)_\mu(x),c_R(x)]
   -z_2\Bar{d}(t,x)c_R(x)^2\bigl]=0.
\label{eq:(4.47)}
\end{align}
This shows that $z_1=z_2=0$ and, by~Eq.~\eqref{eq:(4.46)},
$\Tilde{\Tilde{\mathit{\Gamma}}}_{D+1}^{(\ell+1)\text{div}}=0$.

Thus, going back to~Eq.~\eqref{eq:(4.44)}, we conclude that the most general
form of $\Tilde{\mathit{\Gamma}}^{(\ell+1)\text{div}}$ is
\begin{align}
   \Tilde{\mathit{\Gamma}}^{(\ell+1)\text{div}}
   &=-\frac{1}{2\mu^{2\varepsilon}g^2}
   \int d^Dx\,\tr\left[x_1[F_R]_{\mu\nu}(x)[F_R]_{\mu\nu}(x)\right]
\notag\\
   &\qquad{}
   -2\Tilde{\mathit{\Gamma}}^{(0)}\ast
   \int d^Dx\,\tr\left[y_1(\Tilde{K}_R^A)_\mu(x)(A_R)_\mu(x)
   +y_2K_R^c(x)c_R(x)\right]
\notag\\
   &=-\frac{1}{2\mu^{2\varepsilon}g^2}
   \int d^Dx\,\tr\Bigl\{
   (x_1-2y_1)
   \left[\partial_\mu(A_R)_\nu-\partial_\nu(A_R)_\mu\right]^2
\notag\\
   &\qquad\qquad\qquad\qquad\qquad{}
   +2(x_1-3y_1)
   \left[\partial_\mu(A_R)_\nu(x)-\partial_\nu(A_R)_\mu(x)\right]
\notag\\
   &\qquad\qquad\qquad\qquad\qquad\qquad\qquad\qquad\qquad{}
   \times[(A_R)_\mu(x),(A_R)_\nu(x)]
\notag\\
   &\qquad\qquad\qquad\qquad\qquad\qquad{}
   +(x_1-4y_1)
   [(A_R)_\mu(x),(A_R)_\nu(x)]^2
   \Bigr\}
\notag\\
   &\qquad{}
   -2\int d^Dx\,\tr\Bigl[
   -(y_1+y_2)(\Tilde{K}_R^A)_\mu(x)\partial_\mu c_R(x)
   -y_2(\Tilde{K}_R^A)_\mu(x)[(A_R)_\mu(x),c_R(x)]
\notag\\
   &\qquad\qquad\qquad\qquad\qquad{}
   +y_2K_R^c(x)c_R(x)^2\Bigr]
\notag\\
   &\qquad\qquad{}
   -2\int d^Dx\,\tr\left[
   y_1L_\mu(0,x)(A_R)_\mu(x)
   -y_2\Bar{d}(0,x)c_R(x)\right],
\label{eq:(4.48)}
\end{align}
with three constants $x_1$, $y_1$, and~$y_2$.

Then, it is straightforward to see that if one chooses the $(\ell+1)$-th loop
order coefficients in $Z=\sum_{\ell'=0}^{\ell+1}Z^{(\ell')}$,
$Z_3=\sum_{\ell'=0}^{\ell+1}Z_3^{(\ell')}$, and
$\Tilde{Z}_3=\sum_{\ell'=0}^{\ell+1}\Tilde{Z}_3^{(\ell')}$, as\footnote{Note that
$Z^{(0)}=Z_3^{(0)}=\Tilde{Z}_3^{(0)}=1$.}
\begin{equation}
   Z^{(\ell+1)}=x_1,\qquad
   Z_3^{(\ell+1)}=-x_1+2y_1,\qquad
   \Tilde{Z}_3^{(\ell+1)}=-y_1-y_2,
\label{eq:(4.49)}
\end{equation}
the counterterm in~Eq.~\eqref{eq:(4.25)} precisely cancels the $(\ell+1)$-th
order divergence~\eqref{eq:(4.48)}. In particular, the boundary
counterterm~\eqref{eq:(4.26)} cancels the last line of~Eq.~\eqref{eq:(4.48)}.
This completes the mathematical induction: We have shown that an appropriate
order by order choice of renormalization constants
in~Eqs.~\eqref{eq:(4.15)}--\eqref{eq:(4.19)} makes the generating
functional of the 1PI diagrams---thus all correlation functions of
renormalized fields---finite. In particular, the flowed gauge field does not
require wave function renormalization.

\subsection{Renormalization of fermion fields}
\label{sec:4.5}
One can include the fermion fields into the above argument. Corresponding to
the BRS transformations in~Eqs.~\eqref{eq:(2.6)}, \eqref{eq:(3.8)},
and~\eqref{eq:(3.9)}, the action~$S_K$~\eqref{eq:(4.3)} has additional terms,
\begin{align}
   S_K&=\dotsb
   +\int d^Dx\left[
   \Bar{K}^\psi(x)c(x)\psi(x)+\Bar{\psi}(x)c(x)K^\psi(x)
   \right]
\notag\\
   &\qquad\qquad{}
   +\fint_0^\infty dt\,\int d^Dx\left[
   \Bar{K}^\chi(t,x)d(t,x)\chi(t,x)+\Bar{\chi}(t,x)d(t,x)K^\chi(t,x)
   \right]
\notag\\
   &\qquad\qquad\qquad{}
   +\int_0^\infty dt\,\int d^Dx\left[
   \Bar{K}^\lambda(t,x)d(t,x)\lambda(t,x)+\Bar{\lambda}(t,x)d(t,x)K^\lambda(t,x)
   \right].
\label{eq:(4.50)}
\end{align}
The renormalization constants are introduced as
\begin{align}
   m_0&=Z_m^{-1}m_R,&&
\label{eq:(4.51)}
\\
   \psi&=Z_\psi^{-1/2}\psi_R,&
   \Bar{\psi}&=Z_\psi^{-1/2}\Bar{\psi}_R,
\label{eq:(4.52)}
\\
   \Bar{K}^\psi&=Z_\psi^{1/2}\Bar{K}_R^\psi,&
   K^{\psi}&=Z_\psi^{1/2}K_R^\psi,
\label{eq:(4.53)}
\end{align}
and
\begin{align}
   \Bar{\lambda}&=Z_\chi^{1/2}\Bar{\lambda}_R,&
   \lambda&=Z_\chi^{1/2}\lambda_R,
\label{eq:(4.54)}
\\
   K^\lambda&=Z_\chi^{-1/2}K_R^\lambda,&
   \Bar{K}^\lambda&=Z_\chi^{-1/2}\Bar{K}_R^\lambda,
\label{eq:(4.55)}
\end{align}
and
\begin{align}
   \chi&=Z_\chi^{-1/2}\chi_R,&
   \Bar{\chi}&=Z_\chi^{-1/2}\Bar{\chi}_R,
\label{eq:(4.56)}
\\
   \Bar{K}^\chi&=Z_\chi^{1/2}\Bar{K}_R^\chi,&
   K^\chi&=Z_\chi^{1/2}K_R^\chi.
\label{eq:(4.57)}
\end{align}
The derivative operator~$\Tilde{\mathit{\Gamma}}^{(0)}\ast$ has additional
terms,
\begin{align}
   \Tilde{\mathit{\Gamma}}^{(0)}\ast
   &=\dotsb
   -\int d^Dx\,
   \Biggl[
   \frac{\delta\Tilde{\mathit{\Gamma}}^{(0)}}{\delta\psi_R(x)}
   \frac{\delta}{\delta\Bar{K}_R^\psi(x)}
   +\frac{\delta\Tilde{\mathit{\Gamma}}^{(0)}}{\delta\Bar{K}_R^\psi(x)}
   \frac{\delta}{\delta\psi_R(x)}
\notag\\
   &\qquad\qquad\qquad\qquad{}
   +\frac{\delta\Tilde{\mathit{\Gamma}}^{(0)}}{\delta K_R^\psi(x)}
   \frac{\delta}{\delta\Bar{\psi}_R(x)}
   +\frac{\delta\Tilde{\mathit{\Gamma}}^{(0)}}{\delta\Bar{\psi}_R(x)}
   \frac{\delta}{\delta K_R^\psi(x)}
   \Biggr]
\notag\\
   &\qquad\qquad{}
   -\epsilon\int d^Dx\,
   \Biggl[
   \frac{\delta\Tilde{\mathit{\Gamma}}^{(0)}}{\delta\lambda_R(0,x)}
   \frac{\delta}{\delta\Bar{K}_R^\lambda(0,x)}
   +\frac{\delta\Tilde{\mathit{\Gamma}}^{(0)}}{\delta\Bar{K}_R^\lambda(0,x)}
   \frac{\delta}{\delta\lambda_R(0,x)}
\notag\\
   &\qquad\qquad\qquad\qquad\qquad{}
   +\frac{\delta\Tilde{\mathit{\Gamma}}^{(0)}}{\delta K_R^\lambda(0,x)}
   \frac{\delta}{\delta\Bar{\lambda}_R(0,x)}
   +\frac{\delta\Tilde{\mathit{\Gamma}}^{(0)}}{\delta\Bar{\lambda}_R(0,x)}
   \frac{\delta}{\delta K_R^\lambda(0,x)}
   \Biggr]
\notag\\
   &\qquad\qquad\qquad\qquad\qquad\qquad{}
   +(\text{derivatives w.r.t.\ field variables at $t\neq0$}).
\label{eq:(4.58)}
\end{align}
Following the same line of argument as the pure Yang--Mills case, the
representation of the possible divergent term, Eq.~\eqref{eq:(4.44)}, is
modified to
\begin{align}
   \Tilde{\mathit{\Gamma}}^{(\ell+1)\text{div}}
   &=\dotsb
   +\int d^Dx\,\left\{
   w_1\Bar{\psi}_R(x)\gamma_\mu\left[\partial_\mu+(A_R)_\mu(x)
   \right]\psi_R(x)
   +w_2m_R\Bar{\psi}_R(x)\psi_R(x)\right\}
\notag\\
   &\qquad\qquad{}
   +\Tilde{\mathit{\Gamma}}^{(0)}\ast
   \int d^Dx\,w_3
   \left[\Bar{K}_R^\psi(x)\psi_R(x)+\Bar{\psi}_R(x)K_R^\psi(x)\right]
\notag\\
   &\qquad\qquad\qquad{}
   +\int d^Dx\,\xi_1
   \left[\Bar{\lambda}_R(0,x)\psi_R(x)+\Bar{\psi}_R(x)\lambda_R(0,x)\right],
\label{eq:(4.59)}
\end{align}
where $w_1$, $w_2$, $w_3$, and~$\xi_1$ are (divergent) constants. This time,
unlike the pure Yang--Mills case~\eqref{eq:(4.44)}, the last
term of~Eq.~\eqref{eq:(4.59)} is manifestly BRS (or gauge) invariant and it is
annihilated by~$\Tilde{\mathit{\Gamma}}^{(0)}\ast$. Thus we cannot conclude
$\xi_1=0$ from the WT
relation~$\Tilde{\mathit{\Gamma}}^{(0)}\ast\Tilde{\mathit{\Gamma}}^{(\ell+1)\text{div}}=0$.\footnote{Recall the argument around~Eq.~\eqref{eq:(4.47)} that
eliminated the possibility of $z_1$ and~$z_2$.}

Eq.~\eqref{eq:(4.59)} then yields
\begin{align}
   &\Tilde{\mathit{\Gamma}}^{(\ell+1)\text{div}}
\notag\\
   &=\dotsb
   +\int d^Dx\,\Bigl\{
   (w_1+2w_3)\Bar{\psi}_R(x)\gamma_\mu\partial_\mu\psi_R(x)
\notag\\
   &\qquad\qquad\qquad\qquad{}
   +(w_1+2w_3-y_1)\Bar{\psi}_R(x)\gamma_\mu(A_R)_\mu(x)\psi_R(x)
\notag\\
   &\qquad\qquad\qquad\qquad\qquad{}
   +(w_2+2w_3)m_R\Bar{\psi}_R(x)\psi_R(x)
\notag\\
   &\qquad\qquad\qquad\qquad\qquad\qquad{}
   +y_2\left[
   \Bar{K}_R^\psi(x)c_R(x)\psi_R(x)+\Bar{\psi}_R(x)c_R(x)K_R^\psi(x)
   \right]
\notag\\
   &\qquad\qquad\qquad\qquad\qquad\qquad\qquad{}
   +(-w_3+\xi_1)\left[
   \Bar{\lambda}_R(0,x)\psi_R(x)+\Bar{\psi}_R(x)\lambda_R(0,x)
   \right]
   \Bigr\}.
\label{eq:(4.60)}
\end{align}
These divergences can be canceled by setting the $(\ell+1)$-th order
renormalization constants as
\begin{equation}
   Z_\psi^{(\ell+1)}=w_1+2w_3,\qquad
   Z_m^{(\ell+1)}=-w_1+w_2,\qquad
   Z_\chi^{(\ell+1)}=w_1+4w_3-2\xi_1.
\label{eq:(4.61)}
\end{equation}
In particular, the last line of~Eq.~\eqref{eq:(4.60)} is canceled by the
boundary counterterm arising from the flow time derivative
in~Eq.~\eqref{eq:(3.2)}. Note that there is no particular reason for
$w_1+4w_3-2\xi_1=0$ (i.e., $Z_\chi=1$) to hold and consequently the Lagrange
multiplier fields for the flowed fermion fields, $\Bar{\lambda}(t,x)$
and~$\lambda(t,x)$, \emph{must be renormalized\/}. This is a crucial difference
from the gauge field for which $L_\mu(t,x)$ is not renormalized.

Finally, the renormalization of the flowed fermion fields $\chi(t,x)$
and~$\Bar{\chi}(t,x)$ in~Eq.~\eqref{eq:(4.56)} being reciprocal to that
of~$\lambda(t,x)$ and~$\Bar{\lambda}(t,x)$ is required, \emph{not\/} to
generate a ``bulk counterterm'' from $S_{\text{fl}}$~\eqref{eq:(3.2)}. As we
emphasized, there is no divergence written as a $(D+1)$-dimensional integral of
a local polynomial of fields. Thus we should not have any $(D+1)$-dimensional
counterterm. This requirement fixes the renormalization factors
in~Eq.~\eqref{eq:(4.56)}.

Thus, unlike the flowed gauge field, the flowed fermion field $\chi(t,x)$
and~$\Bar{\chi}(t,x)$ requires wave function renormalization. As
Eq.~\eqref{eq:(4.61)} indicates, the required renormalization constant~$Z_\chi$
is generally different from that for the original $D$-dimensional fermion
field, $Z_\psi$; an explicit perturbative calculation confirms that this is
actually the case. From this observation, it is expected that this wave
function renormalization of a flowed field persists also for generic matter
fields.

\subsection{Composite operators}
\label{sec:4.6}
Finally, we mention on the finiteness of local products of flowed fields
(i.e., composite operators). In the above, we have shown that a correlation
function of flowed fields
\begin{equation}
   Z_\chi^{(m+l)/2}\left\langle B_{\mu_1}(t_1,x_1)\dotsm B_{\mu_n}(t_n,x_n)
   \chi(s_1,y_1)\dotsm\chi(s_m,y_m)
   \Bar{\chi}(u_1,z_1)\dotsm\Bar{\chi}(u_l,z_l) 
   \right\rangle,
\label{eq:(4.62)}
\end{equation}
is finite under the conventional renormalization in the original $4$-dimensional
gauge theory and an appropriate choice of the wave function renormalization
factor~$Z_\chi$. We can further see that this finiteness persists, even if
some of spacetime points (say, $x_1$ and~$x_2$ in~Eq.~\eqref{eq:(4.62)})
coincide, as far as the corresponding flow times ($t_1$ and~$t_2$ in this
example) are strictly positive. This implies that any local products of flowed
elementary fields (with the factor~$Z_\chi^{1/2}$ for flowed fermion fields) is
finite. Because of this remarkable property, the flow provides a versatile
method to define renormalized composite operators.

This finiteness follows again from the fact that a Feynman loop that consists
only of the flow lines (the heat kernels) identically vanishes. Let us take the
above situation in which two spacetime points $x_1$ and~$x_2$ coincide. In the
loop that contains the points $(t_1,x_1)$ and~$(t_2,x_2=x_1)$, there must exist
at least one propagator. The propagator carries the Gaussian damping
factor~$\sim e^{-(v+w)p^2}$, where $p_\mu$ denotes the loop momentum and $v$
and~$w$ are flow times at end points of the propagator. In this situation, if
$v>0$ or~$w>0$, then the loop integral for the loop absolutely converges
because of the damping factor. If $v=w=0$, on the other hand, one of the
propagators or the heat kernels contained in the loop gives rise to the factor
that behaves like~$\sim e^{-t_1p^2}$ or~$\sim e^{-t_2p^2}$. Thus the loop
integral again converges. This shows that the coincidence $x_1=x_2$ does not
produces new divergence as far as the corresponding flow times are strictly
positive.

\section{Conclusion}
\label{sec:5}
In the present paper, we have given another proof of the renormalizability of
the system defined by the gradient flow and the fermion flow in vector-like
gauge theories, the theorem first proven
in~Ref.~\cite{Luscher:2011bx,Luscher:2013cpa}: Any correlation functions and
composite operators of flowed fields are made finite by the conventional
renormalization of gauge theory and wave function renormalization of the flowed
fermion fields. We believe that our proof is quite accessible if not elegant.
We hope that our proof will be helpful to understand the essential feature of
the flow that underlies its various applications in (lattice) gauge theory.

\section*{Acknowledgments}
The prototype of the present proof was presented  by H.~S.\ in a series of
lectures at the Yukawa Institute for Theoretical Physics, Kyoto University. We
would like to thank the participants, especially,
Sinya Aoki,
Takahiro Doi,
Kengo Kikuchi,
Hiroshi Kunitomo,
and
Shigeki Sugimoto,
for valuable questions and comments which made our proof precise and concise.
The work of H.~S. is supported in part by Grant-in-Aid for Scientific Research
No.~16H03982.

\appendix
\section{The $(D+1)$-dimensional field theory in an enlarged field space}
\label{app:A}
In this Appendix, we show that the BRS transformation of the ghost Lagrange
multiplier~$\Bar{d}$, Eq.~\eqref{eq:(3.10)}, and the BRS invariance of the
$(D+1)$-dimensional action~$S_{\text{fl}}+S_{d\Bar{d}}$, Eq.~\eqref{eq:(3.11)},
can naturally be understood if one enlarges the field space so that the gauge
covariance in a $(D+1)$-dimensional sense manifest. The idea of this enlarged
field space can be found in~Ref.~\cite{Nakazawa:1989jf} in the context of the
stochastic quantization and also in~Ref.~\cite{DelDebbio:2013zaa} in the
context of the gradient flow.

For this, we introduce the gauge potential in the flow-time direction,
$B_t(t,x)$, and set
\begin{align}
   G_{t\mu}(t,x)&\equiv\partial_t B_\mu(t,x)-\partial_\mu B_t(t,x)
   +[B_t(t,x),B_\mu(t,x)],
\label{eq:(A1)}
\\
   D_t&\equiv\partial_t+B_t,
\label{eq:(A2)}
\\
   \overleftarrow{D}_t&\equiv\overleftarrow{\partial}_t-B_t,
\label{eq:(A3)}
\end{align}
and consider a $(D+1)$-dimensional action
\begin{align}
   \Hat{S}_{\text{fl}}
   &=-2\int_0^\infty dt\int d^Dx\,
   \tr\left\{
   L_\mu(t,x)\left[G_{t\mu}(t,x)-D_\nu G_{\nu\mu}(t,x)\right]
   \right\}
\notag\\
   &\qquad{}
   +\int_0^\infty dt\,\int d^Dx\,
   \left[
   \Bar{\lambda}(t,x)\left(D_t-\Delta\right)\chi(t,x)
   +\Bar{\chi}(t,x)
   \left(\overleftarrow{D}_t-\overleftarrow{\Delta}\right)\lambda(t,x)
   \right].
\label{eq:(A4)}
\end{align}
We regard the new gauge potential at~$t=0$, $B_t(t=0,x)$, as an independent
integration variable.

The above $(D+1)$-dimensional action~$\Hat{S}_{\text{fl}}$ is manifestly
invariant under the gauge transformation in the $(D+1)$-dimensional
sense.\footnote{Throughout this appendix, the flow time is assumed to be
continuous. We may discretize the flow time while keeping the gauge invariance
by using lattice regularization as in~Ref.~\cite{Luscher:2013cpa}.}
Correspondingly,
$\Hat{S}_{\text{fl}}$ is invariant under the following $(D+1)$-dimensional BRS
transformation:
\begin{align}
   \Hat{\delta}B_\mu(t,x)&=D_\mu d(t,x),&&
\label{eq:(A5)}
\\
   \Hat{\delta}B_t(t,x)&=D_t d(t,x),&&
\label{eq:(A6)}\\
   \Hat{\delta}L_\mu(t,x)&=[L_\mu(t,x),d(t,x)],&&
\label{eq:(A7)}
\\
   \Hat{\delta}\chi(t,x)&=-d(t,x)\chi(t,x),&
   \Hat{\delta}\Bar{\chi}(t,x)&=-\Bar{\chi}(t,x)d(t,x),
\label{eq:(A8)}
\\
   \Hat{\delta}\lambda(t,x)&=-d(t,x)\lambda(t,x),&
   \Hat{\delta}\Bar{\lambda}(t,x)&=-\Bar{\lambda}(t,x)d(t,x),
\label{eq:(A9)}
\\
   \Hat{\delta}d(t,x)&=-d(t,x)^2,
\label{eq:(A10)}
\end{align}
where $d(t,x)$ is a $(D+1)$-dimensional ghost field. As usual, the above
transformation is nilpotent by construction, $\Hat{\delta}^2=0$.

Next, we introduce a $(D+1)$-dimensional ``gauge fixing term''
$\Hat{S}_{\text{GF}}$ and a corresponding ``ghost-anti-ghost term''
$\Hat{S}_{d\Bar{d}}$ by
\begin{equation}
   \Hat{S}_{\text{GF}}+\Hat{S}_{d\Bar{d}}\equiv
   \Hat{\delta}(-2)\int_0^\infty dt\int d^Dx\,
   \tr\left\{
   -\Bar{d}(t,x)\left[B_t(t,x)-\alpha_0\partial_\mu B_\mu(t,x)\right]\right\}.
\label{eq:(A11)}
\end{equation}
In this expression, $\Bar{d}(t,x)$ is a $(D+1)$-dimensional analogue of the
anti-ghost and its BRS transformation is defined by
\begin{equation}
   \Hat{\delta}\Bar{d}(t,x)=C(t,x),\qquad\Hat{\delta}C(t,x)=0,
\label{eq:(A12)}
\end{equation}
where $C(t,x)=C^a(t,x)T^a$ is a $(D+1)$-dimensional analogue of the
Nakanishi--Lautrup field. Clearly, the nilpotency $\Hat{\delta}^2=0$ is
preserved with this definition. Thus, by construction, the above action whose
explicit form is given by
\begin{align}
   \Hat{S}_{\text{GF}}+\Hat{S}_{d\Bar{d}}
   &=
   -2\int_0^\infty dt\int d^Dx\,
   \tr\left\{
   -C(t,x)\left[B_t(t,x)-\alpha_0\partial_\mu B_\mu(t,x)\right]\right\}
\notag\\
   &\qquad{}-2\int_0^\infty dt\int d^Dx\,
   \tr\left\{
   \Bar{d}(t,x)\left[D_t d(t,x)-\alpha_0\partial_\mu D_\mu d(t,x)\right]
   \right\},
\label{eq:(A13)}
\end{align}
is manifestly BRS invariant:
\begin{equation}
   \Hat{\delta}\left(\Hat{S}_{\text{GF}}+\Hat{S}_{d\Bar{d}}\right)=0.
\label{eq:(A14)}
\end{equation}

So far, the nilpotency and the BRS invariance of the action are manifest. We
will now consider a reduced field space in which $B_t$ and~$C$ are eliminated
by using the equations of motion (we denote them EOM in what follows):
\begin{align}
   B_t(t,x)&=\alpha_0\partial_\mu B_\mu(t,x),
\notag\\
   &\Leftrightarrow\frac{\delta}{\delta C^a(t,x)}
   \left(\Hat{S}_{\text{fl}}+\Hat{S}_{\text{GF}}+\Hat{S}_{d\Bar{d}}\right)=0,
\label{eq:(A15)}
\\
   C(t,x)&=D_\mu L_\mu(t,x)-\left\{d(t,x),\Bar{d}(t,x)\right\}
   +\Bar{\lambda}(t,x)T^a\chi(t,x)T^a
   -\Bar{\chi}(t,x)T^a\lambda(t,x)T^a,
\notag\\
   &\Leftrightarrow\frac{\delta}{\delta B_t^a(t,x)}
   \left(\Hat{S}_{\text{fl}}+\Hat{S}_{\text{GF}}+\Hat{S}_{d\Bar{d}}\right)=0.
\label{eq:(A16)}
\end{align}
Let us denote the BRS transformation after the elimination of~$B_t$ and~$C$
as~$\delta$. This is given by Eqs.~\eqref{eq:(3.6)}--\eqref{eq:(3.9)} in the
text and
\begin{align}
   \delta\Bar{d}(t,x)&=\left.C(t,x)\right|_{\text{EOM}}
\notag\\
   &=D_\mu L_\mu(t,x)-\left\{d(t,x),\Bar{d}(t,x)\right\}
   +\Bar{\lambda}(t,x)T^a\chi(t,x)T^a
   -\Bar{\chi}(t,x)T^a\lambda(t,x)T^a.
\label{eq:(A17)}
\end{align}
Interestingly, even after this elimination of~$B_t$ and~$C$, the BRS
transformation remains nilpotent, $\delta^2=0$. The only non-trivial relation
for this is
\begin{align}
   \delta^2\Bar{d}(t,x)
   &=\delta\left[
   D_\mu L_\mu(t,x)-\left\{d(t,x),\Bar{d}(t,x)\right\}
   +\Bar{\lambda}(t,x)T^a\chi(t,x)T^a
   -\Bar{\chi}(t,x)T^a\lambda(t,x)T^a\right]
\notag\\
   &=0.
\label{eq:(A18)}
\end{align}
This relation may directly be confirmed (cf.~Ref~\cite{Luscher:2011bx}).
Instead, one can see this by noting the commutation relation,
\begin{equation}
   \left[\Hat{\delta},\frac{\delta}{\delta B_t^a(t,x)}\right]
   =-f^{abc}d^b(t,x)\frac{\delta}{\delta B_t^c(t,x)},
\label{eq:(A19)}
\end{equation}
which immediately follows from the definition of~$\Hat{\delta}$. By applying
this to~$\Hat{S}_{\text{fl}}+\Hat{S}_{\text{GF}}+\Hat{S}_{d\Bar{d}}$ and using
$\Hat{\delta}(\Hat{S}_{\text{fl}}+\Hat{S}_{\text{GF}}+\Hat{S}_{d\Bar{d}})=0$, we
have
\begin{equation}
   \Hat{\delta}\frac{\delta}{\delta B_t^a(t,x)}
   \left(\Hat{S}_{\text{fl}}+\Hat{S}_{\text{GF}}+\Hat{S}_{d\Bar{d}}\right)
   =
   -f^{abc}d^b(t,x)\frac{\delta}{\delta B_t^c(t,x)}
   \left(\Hat{S}_{\text{fl}}+\Hat{S}_{\text{GF}}+\Hat{S}_{d\Bar{d}}\right).
\label{eq:(A20)}
\end{equation}
Since the right-hand side vanishes under the EOM,
\begin{equation}
   \left.\left\{\Hat{\delta}\left[
   D_\mu L_\mu^a(t,x)-\left\{d(t,x),\Bar{d}(t,x)\right\}^a
   +\Bar{\lambda}(t,x)T^a\chi(t,x)-\Bar{\chi}(t,x)T^a\lambda(t,x)\right]
   \right\}\right|_{\text{EOM}}=0,
\label{eq:(A21)}
\end{equation}
where we have used $\Hat{\delta}C^a(t,x)=0$. Finally, since
$(\Hat{\delta}\Bar{d})|_{\text{EOM}}=\delta\Bar{d}$, we have
\begin{equation}
   \delta\left[
   D_\mu L_\mu^a(t,x)-\left\{d(t,x),\Bar{d}(t,x)\right\}^a
   +\Bar{\lambda}(t,x)T^a\chi(t,x)-\Bar{\chi}(t,x)T^a\lambda(t,x)\right]=0.
\label{eq:(A22)}
\end{equation}
This is the relation~\eqref{eq:(A18)}.

If we integrate over $C(t,x)$, $B_t(t,x)$ is eliminated from the system and we
have the action in the reduced field space~$S_{\text{fl}}+S_{d\Bar{d}}$
in~Eqs.~\eqref{eq:(3.2)} and~\eqref{eq:(3.3)}.  We can see that the reduced
action~$S_{\text{fl}}+S_{d\Bar{d}}$ is invariant under the reduced BRS
transformation~$\delta$. The argument proceeds as follows:
\begin{align}
   &\delta\left(S_{\text{fl}}+S_{d\Bar{d}}\right)
\notag\\
   &=\delta\left[
   \left.
   \left(\Hat{S}_{\text{fl}}+\Hat{S}_{\text{GF}}+\Hat{S}_{d\Bar{d}}\right)
   \right|_{\text{EOM}}\right]
\notag\\
   &=\biggl\{\biggl[
   \int_0^\infty dt\int d^Dx\,
   \alpha_0\partial_\mu D_\mu d^a(t,x)\,\frac{\delta}{\delta B_t^a(t,x)}
\notag\\
   &\qquad\qquad{}
   +\Hat{\delta}
   -\int_0^\infty dt\int d^Dx\,
   D_td^a(t,x)\,\frac{\delta}{\delta B_t^a(t,x)}
   \biggr]\left(\Hat{S}_{\text{fl}}+\Hat{S}_{\text{GF}}+\Hat{S}_{d\Bar{d}}\right)
   \biggr\}\biggr|_{\text{EOM}}
\notag\\
   &=\left.\left[
   \Hat{\delta}
   \left(\Hat{S}_{\text{fl}}+\Hat{S}_{\text{GF}}+\Hat{S}_{d\Bar{d}}\right)
   \right]\right|_{\text{EOM}}
\notag\\
   &=0.
\label{eq:(A23)}
\end{align}
In the second equality, we have noted
\begin{align}
   \delta\left.\left[B_t(t,x)\right]\right|_{\text{EOM}}
   -\left.\left[\Hat{\delta}B_t(t,x)\right]\right|_{\text{EOM}}
   &=\alpha_0\partial_\mu D_\mu d(t,x)-\left.D_td(t,x)\right|_{\text{EOM}},
\label{eq:(A24)}
\\
   \delta\left.(\text{other fields})\right|_{\text{EOM}}
   -\left.\left[\Hat{\delta}(\text{other fields})\right]\right|_{\text{EOM}}
   &=0,
\label{eq:(A25)}
\end{align}
and, in the third equality, we have used the fact that
$\frac{\delta}{\delta B_t^a(t,x)}
(\Hat{S}_{\text{fl}}+\Hat{S}_{\text{GF}}+\Hat{S}_{d\Bar{d}})=0$ under the EOM.

The bottom line of the above argument is
\begin{equation}
   \delta^2=0,\qquad
   \delta\left(S_{\text{fl}}+S_{d\Bar{d}}\right)=0,
\label{eq:(A26)}
\end{equation}
without using any equation of motion in the reduced field space; this is the
property that we employed in the main text.

\section{The WT relation with the discretized flow time}
\label{app:B}
Throughout this appendix, the symbols~$\partial_t$ and $\int dt$ always stand
for the difference operator~\eqref{eq:(3.22)} and the sum~\eqref{eq:(3.21)},
respectively, and do not refer to their continuum counterparts.

When the flow-time derivative is replaced by the difference
operator~\eqref{eq:(3.22)}, the Leibniz rule does not hold and the BRS
transformations on quantities containing~$\partial_t$
in~Eqs.~\eqref{eq:(3.13)}--\eqref{eq:(3.16)} are modified by
$O(\epsilon)$~terms as\footnote{The BRS transformations on elementary fields,
Eqs.~\eqref{eq:(3.6)}--\eqref{eq:(3.9)}, do not contain the flow-time
derivative and they remain nilpotent~$\delta^2=0$ even if the flow time is
discretized. As a consequence, the BRS invariance of the source
term~$S_K$~\eqref{eq:(4.3)} is preserved.}
\begin{align}
   \delta E_\mu(t,x)&=\left[E_\mu(t,x),d(t,x)\right]+D_\mu e(t,x)
   +\epsilon\left[\partial_t B_\mu(t,x),\partial_t d(t,x)\right],
\label{eq:(B1)}
\\
   \delta e(t,x)&=-\left\{e(t,x),d(t,x)\right\}
   -\epsilon\partial_t d(t,x)\partial_t d(t,x),
\label{eq:(B2)}
\\
   \delta f(t,x)&=-d(t,x)f(t,x)-e(t,x)\chi(t,x)
   -\epsilon\partial_t d(t,x)\partial_t\chi(t,x),
\label{eq:(B3)}
\\
   \delta\Bar{f}(t,x)&=-\Bar{f}(t,x)d(t,x)-\Bar{\chi}(t,x)e(t,x)
   -\epsilon\partial_t\Bar{\chi}(t,x)\partial_t d(t,x).
\label{eq:(B4)}
\end{align}
As a consequence, the BRS invariance of the action, Eq.~\eqref{eq:(3.11)}, and
the resultant WT relation~\eqref{eq:(4.5)} are broken by~$O(\epsilon)$ terms.
To be definite, we have
\begin{align}
   \delta\left(S_{\text{fl}}+S_{d\Bar{d}}\right)
   &=-2\int_0^\infty dt\int d^Dx\,\tr\left[L_\mu(t,x)
   \epsilon\left[\partial_t B_\mu(t,x),\partial_t d(t,x)\right]
   \right]
\notag\\
   &\qquad{}
   +\int_0^\infty dt\int d^Dx\,\left[
   \Bar{\lambda}(t,x)\epsilon\partial_t d(t,x)\partial_t\chi(t,x)
   -\epsilon\partial_t\Bar{\chi}(t,x)\partial_t d(t,x)\lambda(t,x)\right]
\notag\\
   &\qquad\qquad{}
   -2\int_0^\infty dt\int d^Dx\,\tr\left[\Bar{d}(t,x)
   \epsilon\partial_t d(t,x)\partial_t d(t,x)\right].
\label{eq:(B5)}
\end{align}
The WT relation~\eqref{eq:(4.5)} is then modified by the expectation value of
this combination. In what follows, we show that the above $O(\epsilon)$ terms
do not contribute in the $\epsilon\to0$ limit and the WT relation assumes the
present form~\eqref{eq:(4.5)}; this point requires careful examination because
some $O(1/\epsilon)$ quantities are hidden in our Feynman rules.\footnote{In
fact, in the proof of the renormalizability of the gradient flow in the 2D
$O(N)$ non-linear sigma model~\cite{Makino:2014sta}, similar seemingly
$O(\epsilon)$ terms survive even in the~$\epsilon\to0$ limit and play an
important role.}

First of all, Eq.~\eqref{eq:(B5)} is an $O(\epsilon)$ quantity and thus its
contribution basically vanishes as~$\epsilon\to0$ unless some element
of~$O(1/\epsilon)$ emerges. The unique place in our Feynman rules where such an
$O(1/\epsilon)$ quantity can appear is the $t=s$ case of
\begin{align}
   \left\langle\partial_t B_\mu^a(t,x)L_\nu^b(s,y)\right\rangle_0
   &=\delta^{ab}\delta(t-s)K_{t-s-\epsilon}^\epsilon(x-y)_{\mu\nu}
   +\delta^{ab}\vartheta(t-s+\epsilon)
   \partial_tK_{t-s-\epsilon}^\epsilon(x-y)_{\mu\nu},
\label{eq:(B6)}
\\
   \left\langle\partial_t\chi(t,x)\Bar{\lambda}(s,y)\right\rangle_0
   &=\delta(t-s)K_{t-s-\epsilon}^\epsilon(x-y;1)
   +\vartheta(t-s+\epsilon)
   \partial_tK_{t-s-\epsilon}^\epsilon(x-y;1),
\label{eq:(B7)}
\\
   \left\langle\lambda(t,x)\partial_s\Bar{\chi}(s,y)\right\rangle_0
   &=\delta(s-t)K_{s-t-\epsilon}^\epsilon(x-y;1)
   +\vartheta(s-t+\epsilon)
   \partial_sK_{s-t-\epsilon}^\epsilon(x-y;1),
\label{eq:(B8)}
\\
   \left\langle\partial_t d^a(t,x)\Bar{d}^b(s,y)\right\rangle_0
   &=\delta^{ab}\delta(t-s)K_{t-s-\epsilon}^\epsilon(x-y;\alpha_0)
   +\delta^{ab}\vartheta(t-s+\epsilon)
   \partial_tK_{t-s-\epsilon}^\epsilon(x-y;\alpha_0),
\label{eq:(B9)}
\end{align}
where the ``delta function'' has been introduced by
\begin{equation}
   \delta(t)\equiv\partial_t\vartheta(t)
   =\begin{cases}
   0,&\text{for $t>0$},\\
   \frac{\displaystyle1}{\displaystyle\epsilon},&\text{for $t=0$},\\
   0,&\text{for $t<0$}.\\
   \end{cases}
\label{eq:(B10)}
\end{equation}
Note that $\delta(0)=1/\epsilon$ and this factor can potentially cancel the
factor~$\epsilon$.

Let us consider the expectation value of the right-hand side
of~Eq.~\eqref{eq:(B5)} under the functional integral~\eqref{eq:(4.4)}. For
example, for the first line of~Eq.~\eqref{eq:(B5)}, we consider a flow line
starting from the factor~$\partial_t B_\mu(t,x)$. The two-point
function~\eqref{eq:(B6)} is attached to this flow line. If this flow line ends
at a certain interaction vertex contained in the
action~$S_{\text{tot}}+S_J+S_K$, since those interaction vertices do not
contain~$\partial_s$, the factor $\delta(0)=1/\epsilon$ does not arise because
of the flow-time integration at the vertex, $\int ds\,\delta(t-s)=1$; the
contribution remains $O(\epsilon)$. The cancellation
$\epsilon\cdot1/\epsilon=1$ can occur only when the flow line goes back
to~Eq.~\eqref{eq:(B5)} itself, i.e., only when Eq.~\eqref{eq:(B5)} is
self-contracted by~Eq.~\eqref{eq:(B6)}. A similar argument applies to other
lines of~Eq.~\eqref{eq:(B5)}. However, the self-contractions are proportional
to the factor $f^{abc}\delta^{ab}$ or~$\tr T^a$ which identically
vanishes.\footnote{With dimensional regularization, the self-contractions
vanish also because $\lim_{\epsilon\to0}K_{-\epsilon}^\epsilon(0)_{\mu\nu}
=\lim_{\epsilon\to0}K_{-\epsilon}^\epsilon(0;\alpha_0)=0$ for a complex~$D$.}
This shows that the WT relation restores the desired form~\eqref{eq:(4.5)}
safely in the $\epsilon\to0$ limit.

\end{document}